\documentclass[preprint,preprintnumbers,showpacs,aps,prd,amssymb]{revtex4-1}

\usepackage{graphicx}
\usepackage{bm}
\usepackage{amsmath}
\usepackage{xcolor}
\def\calC{{\cal C}}

\def\calH{{\cal H}}

\def\calO{{\cal O}}

\def\Bbar{{\bar B}}

\def\cbar{{\bar c}}

\def\hbar{{\bar h}}

\def\SM{{\rm SM}}
\def\NP{{\rm NP}}
\def\GeV{{\rm GeV}}
\def\TeV{{\rm TeV}}
\def\Br{{\rm Br}}

\def\Re{{\rm Re}}
\def\min{{\rm min}}
\def\dof{{\rm d.o.f.}}

\def\Ds{D^{(*)}}
\def\Ks{K^{(*)}}

\def\nn{\nonumber}

\begin{document}
\title{New physics effects on $B\to\Ds\tau\nu$ decays}
\author{Jong-Phil Lee}
\email{jongphil7@gmail.com}
\affiliation{Sang-Huh College, Konkuk University, Seoul 05029, Korea}

\begin{abstract}
We investigate new physics effects on $B\to\Ds\tau\nu$ decays in a general and model-independent way.
The $\chi^2$ fits for fractions of the branching ratios $R(\Ds)$ and other polarization parameters are implemented.
We parameterize the relevant Wilson coefficients with a new physics scale and its power
together with combined fermionic couplings.
Constraints from $B_c\to\tau\nu$ are imposed such that its branching ratio is less than 30\%.
For a moderate range of our parameters we find that the new physics scale goes up to $\lesssim 27~\TeV$
for ordinary new particle contributions.
It turns out that the polarization asymmetry of $\tau$ for $B\to D$ transition can be negative only for
a few combinations of the new physics operators.
We also discuss related processes $B_c\to J/\Psi\tau\nu$ and $\Lambda_b\to\Lambda_c\tau\nu$ decays.
\end{abstract}

\maketitle
%%%%%%%%%%%%%%%%%%%%%%%%%%%%%%%%%%%%%%%%%%%%%%%%%%%%%%
\section{Introduction}
%%%%%%%%%%%%%%%%%%%%%%%%%%%%%%%%%%%%%%%%%%%%%%%%%%%%%%
%
The standard model (SM) is an important cornerstone of today's particle physics.
It has been very successful for many decades since its advent.
On the other hand, we know that there must be some new physics (NP) beyond the SM.
Flavor physics plays a very important role in probing NP.
Recently various decay modes of $B$ showed some anomalies which deviate from the SM predictions.
\par
Among them is the ratio of $R(\Ds)$ in $b\to c$ transition, which is defined by
\begin{equation}
R(\Ds)\equiv\frac{{\rm Br}(B\to \Ds\tau\nu)}{{\rm Br}(B\to \Ds\ell\nu)}~,
\label{RDsDef}
\end{equation}
where $\ell =\mu$ or $e$.
For $b\to s$ transition we have similar ratios $R(\Ks)$.
Recent measurements from LHCb turn out to be consistent with the SM \cite{LHCb2212_52,LHCb2212_53}, 
alleviating a tension between experiment and theory.
Unlike $R(\Ks)$, $R(\Ds)$ involves charged currents and the processes occur at tree level in the SM.
\par
Besides the fraction of the branching ratios there also exist polarization parameters.
The polarization asymmetry of $\tau$ is defined as
\begin{equation}
P_\tau(\Ds) \equiv \frac{\Gamma^{D^{(*)}}_\tau(+)-\Gamma^{D^{(*)}}_\tau(-)}
                                          {\Gamma^{D^{(*)}}_\tau(+)+\Gamma^{D^{(*)}}_\tau(-)}~,
\label{PDsDef}
\end{equation}
where $\Gamma^{\Ds}_\tau(\pm)$ is the decay width corresponding to $(\pm)~\tau$ helicity. 
\par
As for $D^*$ we can define the longitudinal polarization as 
\begin{equation}
F_L(D^*)\equiv\frac{\Gamma(B\to D_L^*\tau\nu)}{\Gamma(B\to D^*\tau\nu)}~.
\label{FLDef}
\end{equation}
\par
The SM predictions for these observables are \cite{DAlise2403,Iguro2405}
\begin{eqnarray}
R(D)_\SM &=& 0.290\pm0.003~, \label{RDSM} \\
R(D^*)_\SM &=& 0.248\pm0.001~, \label{RDsSM} \\
P_\tau(D)_\SM &=& 0.331\pm0.004~, \label{PDSM} \\
P_\tau(D^*)_\SM &=& -0.497\pm0.007~,\label{PDsSM} \\
F_L(D^*)_\SM &=& 0.464\pm0.003~. \label{FLSM}
\end{eqnarray}
\par
Experimentally, there are some tensions compared with the SM predictions.
Very recent data from the LHCb tell that \cite{LHCb24}
\begin{equation}
R(D) = 0.249\pm0.043\pm0.047~, ~~~
R(D^*) = 0.402\pm0.081\pm0.085~.
\end{equation}
The polarization asymmetry was measured to be \cite{Belle1612,Belle1709}
\begin{equation}
P_\tau(D^*) = -0.38\pm 0.51^{+0.21}_{-0.16}~.
\end{equation}
Recent measurement of the longitudinal $D^*$ polarization by the LHCb is \cite{LHCb2311}
\begin{equation}
F_L(D^*) = 0.43\pm0.06\pm0.03~.
\end{equation}
The value is closer to the SM prediction of Eq.\ (\ref{FLSM}) than the previous measurements
$F_L(D^*) = 0.60\pm0.08\pm0.04$ by Belle \cite{Abdesselam}.
But other observables, especially $R(D^*)$, strongly imply that there might be NP to affect the 
$B\to\Ds\tau\nu$ decay.
\par
There have been many NP scenarios suggested to explain the $B$ anomalies,
including leptoquarks (LQ) \cite{Sakaki1309,Freytsis1506,Bauer1511,Fajfer1511,Becirevic2404}, 
$W'$ \cite{Datta1412,Greljo1506,Boucenna1608,Carena1809}, 
charged Higgs \cite{Crivellin1206,Fajfer1206,Celis1210}, 
or unparticles \cite{JPL2012}.
All of these NP effects occur via tree-level contributions with charged currents.
\par
In this paper we investigate the NP effects in a more general way. 
We parametrized the NP contributions as new Wilson coefficients 
$C_j^\NP\sim A_j(v/M_\NP)^\alpha\Big[\alpha_s(M_\NP)\Big]^{\gamma_j/2\beta_0^{(n_f)}}$ 
where $M_\NP$ is the NP scale, $v$ is the SM vacuum expectation value,
$\gamma_j$ is the anomalous dimension, 
and $\beta_0^{(n_f)}=11-2n_f/3$ with $n_f$ being the number of quark flavors.
Terms with $\alpha_s(M_\NP)$ reflect the renormalization group (RG) running effects.
We allow $\alpha$ to be a free parameter inspired by the unparticle scenario.
The case where $\alpha=2$ corresponds to ordinary new particle contributions at tree level
such as LQ, $W'$, or charged Higgs, etc.
Other details of NP models are encapsulated in the coefficients $A_j$.
In this framework the interplay between the fermionic coupling and the new mediator 
for NP contributions can be easily seen.
For some specific NP models if the allowed range of the relevant couplings and 
the mass scale of the new mediating (un)particle were known from other processes,
then one could check directly the compatibility of the model with $B\to\Ds\tau\nu$ 
with the help of our analysis.
The approach was applied to $R(\Ks)$ previously \cite{JPL2110,JPL2208}.
One can examine general features of NP effects on $B\to\Ds\tau\nu$ decays in model-independent ways.
\par
Closely related observables are $R(J/\Psi)$ and $R(\Lambda_c)$.
They are defined by
\begin{eqnarray}
R(J/\Psi) &=& \frac{\Br(B_c\to J/\Psi\tau\nu)}{\Br(B_c\to J/\Psi\mu\nu)}~,\label{RJPDef}\\
R(\Lambda_c) &=& \frac{\Br(\Lambda_c\to\tau\nu)}{\Br(\Lambda_c\to\ell\nu)}~.\label{RLdcDef}
\end{eqnarray}
Especially it is known that $R(\Lambda_c)$ satisfies a sum rule with respect to $R(\Ds)$ as
\cite{Blanke1811,Blanke1905,Fedele2211}
\begin{equation}
\frac{R(\Lambda_c)}{R(\Lambda_c)_\SM} \simeq 
0.280\frac{R(D)}{R(D)_\SM} + 0.720\frac{R(D^*)}{R(D^*)_\SM}~.
\label{sumrule}
\end{equation}
The SM predicts that $R(J/\Psi)_\SM = 0.258\pm0.004$ \cite{LQCD2007}
and $R(\Lambda_c)_\SM = 0.324\pm0.004$ \cite{Bernlochner1808,Bernlochner1812}.
Recent experimental data reveal some tension with the SM results as
$R(J/\Psi) = 0.61\pm 0.18 $ \cite{Iguro2405} and 
$R(\Lambda_c) = |0.04/V_{cb}|^2(0.285\pm0.073)$ \cite{Bernlochner2206}.
NP effects, if exist, seems to enhance $R(J/\Psi)$ while shrink $R(\Lambda_c)$.
Still there exist large experimental uncertainties for both of them and
we will not take $R(J/\Psi)$ and $R(\Lambda_c)$ as fitting data.
Instead we provide outlooks of NP effects allowed by $B\to\Ds$ transitions on them.
\par
The branching ratio of $\Br(B_c\to\tau\nu)$ puts constraints on NP contributions.
We require a moderate bound of the branching ratio as $\Br(B_c\to\tau\nu)<0.3$ \cite{Alonso1611}.
As will be seen later, most of allowed branching ratios of our analysis are safely below the bound.
\par
The paper is organized as follows.
In Sec.\ II we give the setup for our analysis.
Relevant observables are described by Wilson coefficients with our new parametrization.
Our results and related discussions will appear in Sec.\ III.
Section IV provides conclusions and outlooks.
%
%%%%%%%%%%%%%%%%%%%%%%%%%%%%%%%%%%%%%%%%%%%%%%%%%%%%%%
\section{Setup}
%%%%%%%%%%%%%%%%%%%%%%%%%%%%%%%%%%%%%%%%%%%%%%%%%%%%%%
%
Let's start with the $b\to c$ transition.
The effective Hamiltonian for $b\to c\ell\nu$ is 
\begin{equation}
\calH_{\rm eff}
=\frac{4G_F}{\sqrt{2}}V_{cb}\sum_{\ell=\mu,\tau}\left[
(1+C_{VL}^{\ell})\calO_{VL}^{\ell} + C_{SL}^{\ell}\calO_{SL}^{\ell} + C_{SR}^{\ell}\calO_{SR}^{\ell}
+ C_T^\ell\calO_T^\ell
\right]~,
\label{Heff}
\end{equation}
where the operators $\calO_{V,S}^\ell$ are defined by
\begin{eqnarray}
\calO_{VL}^{\ell} &=& \left(\cbar_L\gamma^\mu b_L\right)\left({\bar\ell}_L\gamma_\mu\nu_{\ell L}\right)~,\\
\calO_{SL}^{\ell} &=& \left(\cbar_R b_L\right) \left({\bar\ell}_R\nu_{\ell L}\right)~,\\
\calO_{SR}^{\ell} &=& \left(\cbar_L b_R\right) \left({\bar\ell}_R\nu_{\ell L}\right)~,\\
\calO_T^\ell &=& (\cbar_R\sigma^{\mu\nu}b_L)({\bar\ell}_R\sigma_{\mu\nu}\nu_{\ell L})~.
\end{eqnarray}
In our analysis the operator 
$\calO_{VR}^\ell=(\cbar_R\gamma^\mu b_R)({\bar\ell}_L\gamma_\mu\nu_{\ell L})$
is neglected for simplicity.
We assume that NP effects appear only in the $\tau$ sector, 
so from now on we omit the superscript $\ell$ of the operators and the Wilson coefficients in Eq.\ (\ref{Heff}).

\par
The observables for $B\to\Ds\tau\nu_\tau$ decays can be described numerically in terms of 
NP Wilson coefficients.
We adopt recently updated formulae for the observables as follows (at $\mu=\mu_b=4.18~\GeV$ scale) \cite{Iguro2405,Blanke1811,Aoki,Bernlochner}
\begin{eqnarray}
\label{RDs}
\frac{R(D)}{R(D)_\SM} &=&
|1+C_{VL}|^2 + 1.01|C_{SL}+C_{SR}|^2 + 0.84|C_T|^2 \nn\\
&&
+ 1.49\Re\left[(1+C_{VL})(C_{SL}+C_{SR})^*\right]
+ 1.08\Re\left[(1+C_{VL})C_T^*\right] ~,\label{RDnum}\\
\frac{R(D^*)}{R(D^*)_\SM} &=& 
|1+C_{VL}|^2 + 0.04|C_{SL}-C_{SR}|^2 + 16.0|C_T|^2 \nn\\
&&
- 0.11\Re\left[(1+C_{VL})(C_{SL}-C_{SR})^*\right]
- 5.17\Re\left[(1+C_{VL})C_T^*\right] ~,\label{RDsnum}
\end{eqnarray}
and
\begin{eqnarray}
\frac{P_\tau(D)}{P_\tau(D)_\SM} &=& 
\left[\frac{R(D)}{R(D)_\SM}\right]^{-1}\Big\{
|1+C_ {VL}|^2 + 3.04|C_{SL}+C_{SR}|^2 + 0.17|C_T|^2 \label{PtD} \nn\\
&& 
+ 4.50\Re\left[(1+C_{VL})(C_{SL}+C_{SR})^*\right]
- 1.09\Re\left(1+C_{VL})C_T^*\right] 
 \Big\}~,\label{PDnum}\\
\frac{P_\tau(D^*)}{P_\tau(D^*)_\SM} &=& 
\left[\frac{R(D^*)}{R(D^*)_\SM}\right]^{-1}\Big\{
|1+C_ {VL}|^2 - 0.07|C_{SL}-C_{SR}|^2 - 1.85|C_T|^2\nn\\
&& 
+ 0.23\Re\left[(1+C_{VL})(C_{SL}-C_{SR})^*\right] 
- 3.47\Re\left[(1+C_{VL})C_T^*\right]
 \Big\}~,\label{PDsnum}\\
\frac{F_L(D^*)}{F_L(D^*)_\SM} &=&
\left[\frac{R(D^*)}{R(D^*)_\SM}\right]^{-1}\Big\{
|1+C_ {VL}|^2 + 0.08|C_{SL}-C_{SR}|^2 + 6.90|C_T|^2\nn\\
&& 
- 0.25\Re\left[(1+C_{VL})(C_{SL}-C_{SR})^*\right] 
- 4.30\Re\left[(1+C_{VL})C_T^*\right]
 \Big\}~,
\label{FLnum}
\end{eqnarray}
%
%-----------------------------------------------------------------------------------------------------------------------------------------
\par
As for $B_c\to J/\Psi\tau\nu$ and $\Lambda_b\to\Lambda_c\tau\nu$ decays, we have \cite{Iguro2405}
\begin{eqnarray}
\frac{R(J/\Psi)}{R(J/\Psi)_\SM} &=& 
|1+C_{VL}|^2 + 0.04|C_{SL}-C_{SR}|^2
-0.10\Re\left[(1+C_{VL})(C_{SL}-C_{SR})^*\right] \nn\\
&&
-5.39\Re\left[(1+C_{VL})C_T^*\right] + 14.7|C_T|^2~, \label{RJPnum} \\
\frac{R(\Lambda_c)}{R(\Lambda_c)_\SM} &=&
|1+C_{VL}|^2 + 0.32\left(|C_{SL}|^2+|C_{SR}|^2\right)
+0.52\Re\left[C_{SL}C_{SR}^*\right] \nn\\
&&
+0.50\Re\left[(1+C_{VL})C_{SR}^*\right] 
+0.33\Re\left[(1+C_{VL})C_{SL}^*\right] \nn\\
&&
-3.11\Re\left[(1+C_{VL})C_T^*\right] + 10.4|C_T|^2~.
\label{RLdcnum}
\end{eqnarray}
\par
Now the Wilson coefficients can be written as \cite{Sakaki1309}
\begin{equation}
C_j(\mu_b) \equiv 
A_j\left(\frac{v}{M_\NP}\right)^\alpha
\left[\frac{\alpha_s(m_t)}{\alpha_s(\mu_b)}\right]^{\frac{\gamma_j}{2\beta_0^{(5)}}}
\left[\frac{\alpha_s(M_\NP)}{\alpha_s(m_t)}\right]^{\frac{\gamma_j}{2\beta_0^{(6)}}}
~, 
\end{equation}
where $A_j$ are some combinations of the relevant fermionic couplings of NP with $j=VL, SL, SR, T$, 
and $M_\NP$ is the NP scale.
The exponent $\alpha$ is taken to be a free parameter.
Terms with $\alpha_s$ are for the RG running effects.
More explicitly, we have \cite{Becirevic2404}
\begin{eqnarray}
C_{VL} &=& A_{VL}\left(\frac{v}{M_\NP}\right)^\alpha~, \\
C_{SL(R)} &=& 
A_{SL(R)}\left(\frac{v}{M_\NP}\right)^\alpha
\left[\frac{\alpha_s(m_t)}{\alpha_s(\mu_b)}\right]^{-\frac{12}{23}}
\left[\frac{\alpha_s(M_\NP)}{\alpha_s(m_t)}\right]^{-\frac{4}{7}}~,\\
C_T &=& 
A_T\left(\frac{v}{M_\NP}\right)^\alpha
\left[\frac{\alpha_s(m_t)}{\alpha_s(\mu_b)}\right]^{\frac{4}{23}}
\left[\frac{\alpha_s(M_\NP)}{\alpha_s(m_t)}\right]^{\frac{4}{21}}~,\\
\end{eqnarray}
where vector current Wilson coefficients do not run.
\par
For example, contributions from the weak singlet scalar leptoquark are \cite{DAlise2403}
\begin{eqnarray}
A_{SL} &=& -\frac{y_{1b\tau}^{LL}(y_{1c\tau}^{RR*})}{2V_{cb}}~,\\
A_{VL} &=& \frac{y_{1b\tau}^{LL}(Vy_1^{LL*})_{c\tau}}{2V_{cb}}~,
\label{S1}
\end{eqnarray}
where $V$ is the CKM matrix and $y_{1ij}$ are the relevant couplings.
Other NP models involve similar contributions.
\par
The branching ratio of $B_c\to\tau\nu$ is also directly related to $C_{ij}$ as \cite{DAlise2403,Iguro2405}
\begin{equation}
\Br(B_c\to\tau\nu)=\Br(B_c\to\tau\nu)_\SM\Big|
1+C_{VL}-4.35(C_{SL}-C_{SR})\Big|^2~,
\end{equation}
where $\Br(B_c\to\tau\nu)_\SM = 0.022$.
We require a moderate bound of $\Br(B_c\to\tau\nu)<0.3$.
%
%
%%%%%%%%%%%%%%%%%%%%%%%%%%%%%%%%%%%%%%%%%%%%%%%%%%%%%%
\section{Results and discussions}
%%%%%%%%%%%%%%%%%%%%%%%%%%%%%%%%%%%%%%%%%%%%%%%%%%%%%%
%
Experimental data used for our fit are summarized in Table \ref{T1}.
%
%-------------------- Table 1---------------------------------------------------
\begin{table}
\begin{tabular}{|c|| cc |}\hline
              & ~$R(D)$ & ~$R(D^*)$   \\\hline
 BABAR & ~$0.440\pm0.058\pm0.042$ & ~$0.332\pm0.024\pm0.018$ \cite{BaBar1} \\
 Belle(2015) & ~$0.375\pm0.064\pm0.026$ & ~$0.293\pm0.038\pm0.015$ \cite{Belle1} \\
 Belle(2016) & ~$-$ & ~$0.302\pm0.030\pm0.011$ \cite{Belle1607} \\
 Belle(2017) & ~$-$ & ~$0.276\pm0.034^{+0.029}_{-0.026}$ \cite{Belle1703} \\
 Belle(2017) & ~$-$ & ~$0.270\pm0.035^{+0.028}_{-0.025}$ \cite{Belle1612,Belle1709} \\
 Belle(2019) & ~$0.307\pm0.037\pm0.016$ & ~$0.283\pm0.018\pm0.014$ \cite{Belle1904}\\
 BelleII(2024) & ~$-$ & ~$0.262^{+0.041+0.035}_{-0.039-0.032}$  \cite{BelleII2024}\\
 LHCb(2015) & ~$-$ & ~$0.336\pm0.027\pm0.030$ \cite{LHCb1} \\
 LHCb(2017) & ~$-$ & ~$0.291\pm0.019\pm0.026\pm0.013$ \cite{LHCb2}  \\
 LHCb(2023) & ~$-$ & ~$0.260\pm0.015\pm0.016\pm0.012$ \cite{LHCb23} \\
 LHCb(2024) & ~$0.249\pm0.043\pm0.047$ & ~$0.402\pm0.081\pm0.085$ \cite{LHCb24} \\ 
 \hline\hline
 & $P_\tau(D^*)$ & $F_L(D^*)$ \\\hline
 Belle(2017) & $-0.38\pm0.51^{+0.21}_{-0.16}$\cite{Belle1612,Belle1709} & $-$ \\
 Belle(2019)  & ~$-$ & ~$0.60\pm0.08\pm0.04$ \cite{Abdesselam} \\
LHCb(2023) & ~$-$ & ~$0.43\pm0.06\pm0.03$ \cite{LHCb2311} \\
 \hline
 \end{tabular}
\caption{Summary of experimental data for $R(\Ds)$, $P_\tau(\Ds)$ and $F_L(D^*)$.
The uncertainties are $\pm$(statistical)$\pm$(systematic).
The correlations between $R(D)$ and $R(D^*)$ for BABAR, Belle(2015),  Belle(2019),
and LHCb(2024) results are
$-0.31$, $-0.50$, $-0.51$, and $-0.39$, respectively \cite{HFAG2019, LHCb24}.}
\label{T1}
\end{table}
%--------------------------------------------------------------------------------------------------------------------------------------
%
The $\chi^2$ is defined by
\begin{equation}
\chi^2\equiv\sum_{i,j}
\left[ \calO_i^{\rm exp}-\calO_i^{\rm th}\right]\calC_{ij}^{-1}
\left[\calO_j^{\rm exp}-\calO_j^{\rm th}\right]~,
\label{chisq}
\end{equation}
where $\calO_i^{\rm exp}$ are the experimental data in Table \ref{T1}
and $\calO_i^{\rm th}$ are the theoretical calculations from Eqs.(\ref{RDnum})-(\ref{FLnum}).
Here $\calC_{ij}$ are the correlation matrix elements.
\par
The fitting parameters scan the range of 
$0\le\alpha\le 7$, $1~\TeV\le M_\NP\le 100~\TeV$, and
$-100\le A_{ij}\le 100$.
Our best-fit values are given in Table \ref{T2}.
We have $\chi^2_\min/\dof = 1.251$.
Note that the best-fit value of $\alpha$ is slightly larger than ordinary value of 2,
but the value of $\chi^2$ at $\alpha=2$ is not so different from $\chi^2_\min$.
% 
%-------------------- Table 2 ---------------------------------------------------
\begin{table}
\begin{tabular}{|cc||cc|}\hline
$~~\alpha$ & $2.278$                            & $~~R(D)$ & $~~~ 0.329~~~$ \\
$~~M_\NP$ & $~~~4.731(\TeV)~~~$  & $~~R(D^*)$ & $~~~ 0.291~~~$  \\
$C_{VL}$ & $~~~ 0.105~~~$          & $~~P_\tau(D)$ & $~~~ 0.258~~~$  \\       
$C_{SL}$ & $~~~ -0.188~~~$        & $~~P_\tau(D^*)$ & $~~~ -0.454~~~$  \\                  
$C_{SR}$ & $~~~ 0.121~~~$         &  $~~F_L(D^*)$ & $~~~ 0.490~~~$  \\
 $C_T$     & $~~~ -0.016~~~$        & $\Br(B_c\to\tau\nu)$ & $~~~ 0.132~~~$ \\
   \hline
\end{tabular}
\caption{Best-fit values.}
\label{T2}
\end{table}
%--------------------------------------------------------------------------------------------------------------
\par
\par
\begin{figure}
\begin{tabular}{c}
\hspace{-1cm}\includegraphics[scale=0.12]{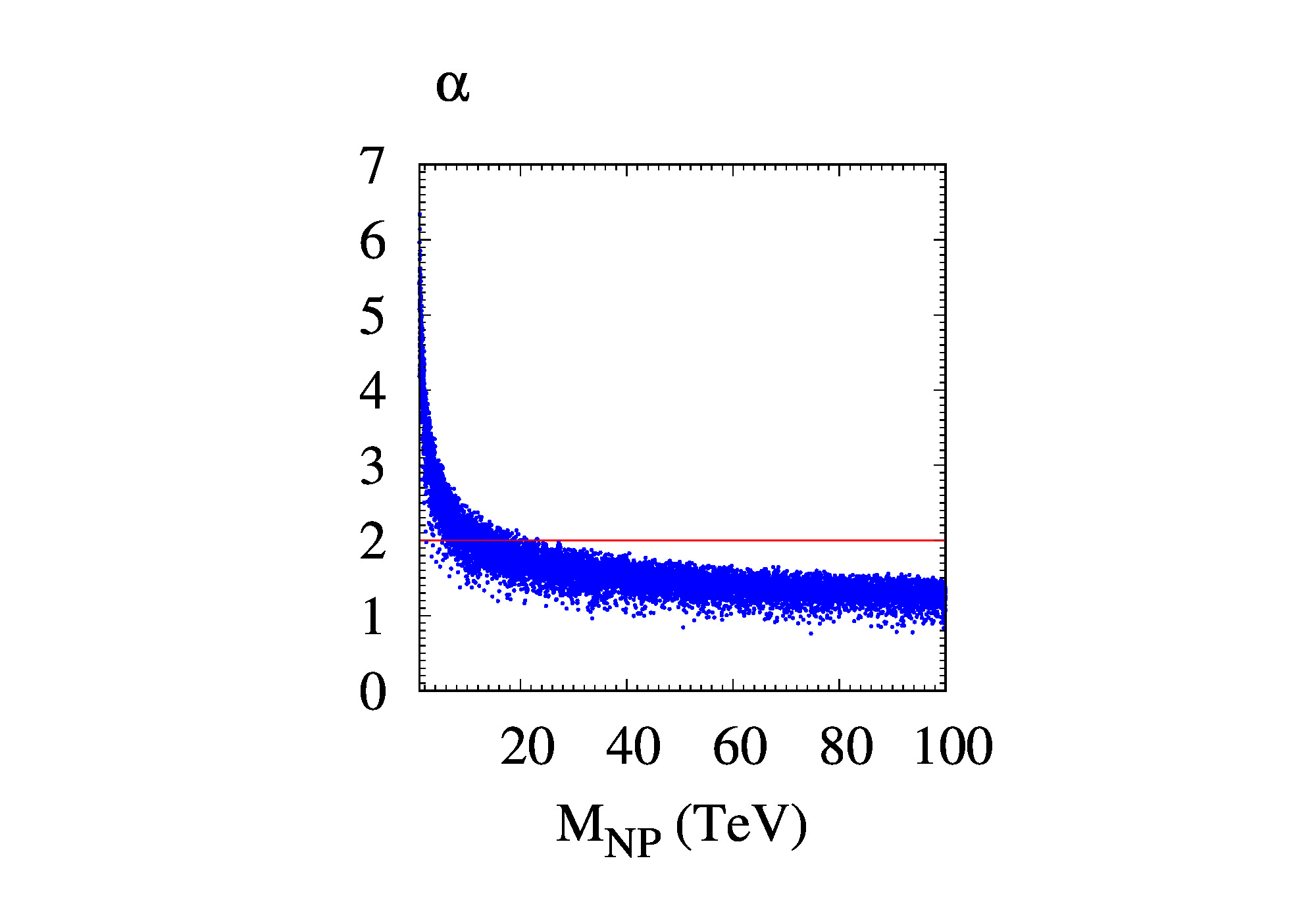}
\end{tabular}
\caption{\label{alp_M} Allowed regions of $\alpha$ vs. $M_{NP}$ at the $2\sigma$ level.}
\end{figure}

Figure \ref{alp_M} shows the allowed region of $\alpha$ vs. $M_\NP$ at the $2\sigma$ level.
%
%----------------- Figure 1 ------------------------------------------------ 
%-------------------------------------------------------------------------------------------------------------
%
For larger values of $M_\NP$ the smaller values of $\alpha$ are allowed.
The horizontal red line corresponds to $\alpha=2$.
In our parameter space of $-100\le A_j \le 100$ the maximum value of $M_\NP$ is about $27~\TeV$
for $\alpha=2$.
Of course the maximum $M_\NP$ could be larger for wider ranges of $A_j$.
For example, if $|A_j|\le 200$ then $M_\NP$ can be as large as 
$\sim 27\times \sqrt{2}\simeq 38~(\TeV)$.
\par
%%%%%%%%%%%%%%%%%%%%%%%%%%%
%
%\textcolor{red}{$<<$ ISSUE3$>>$}
%
%%%%%%%%%%%%%%%%%%%%%%%%%%%
%
Actually direct search of $M_\NP = 27~\TeV$ is out of reach for current accelerators.
To the extent that the logarithmic dependence of the RG running on $M_\NP$ is negligible,
if $A_j$ gets larger (smaller) $k$ times then $M_\NP$ can be as large (small) as $\sqrt{k}$ times.
For example if $k=0.1$ then $M_\NP$ is roughly $M_\NP\lesssim 8.5~\TeV$.
In this way current accelerator could check the validity of NP models with small fermionic couplings.
\par
In the unparticle scenario $\alpha$ can be non-integer and $M_\NP$ is very sensitive to $\alpha$.
For scalar (vector) unparticles $\alpha$ can span $2(4)\le\alpha$ \cite{JPL2012}. 
In case of $\alpha = 4$, we have $M_\NP\lesssim 2.52~ \TeV$.
If there were no unparticle signals below $\sim 2.52~\TeV$ with $|A_j|\le 100$, 
then the vector unparticle scenario would be disfavored.
\par
\begin{figure}
\begin{tabular}{cc}
\hspace{-1cm}\includegraphics[scale=0.12]{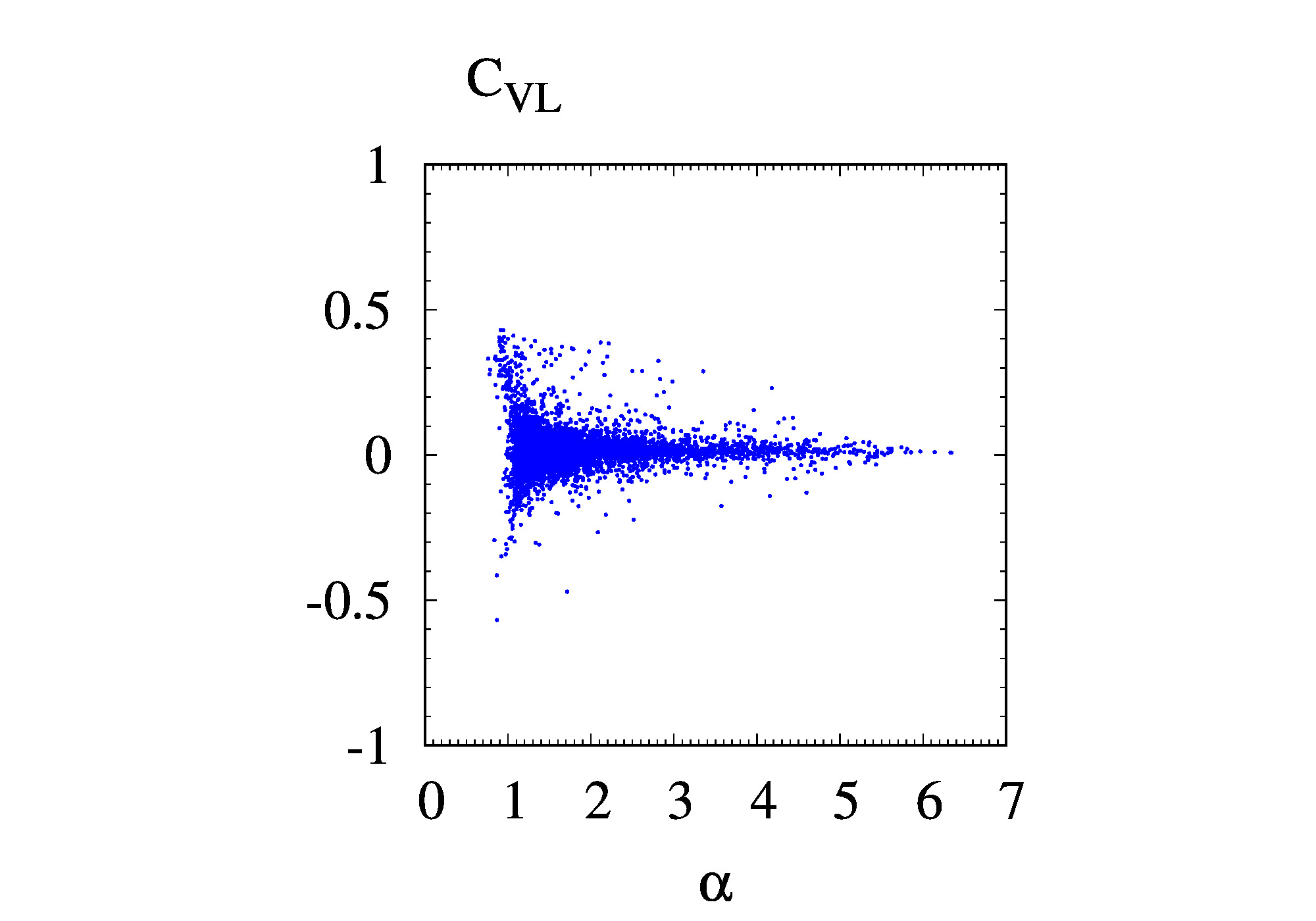}&
\hspace{-1cm}\includegraphics[scale=0.12]{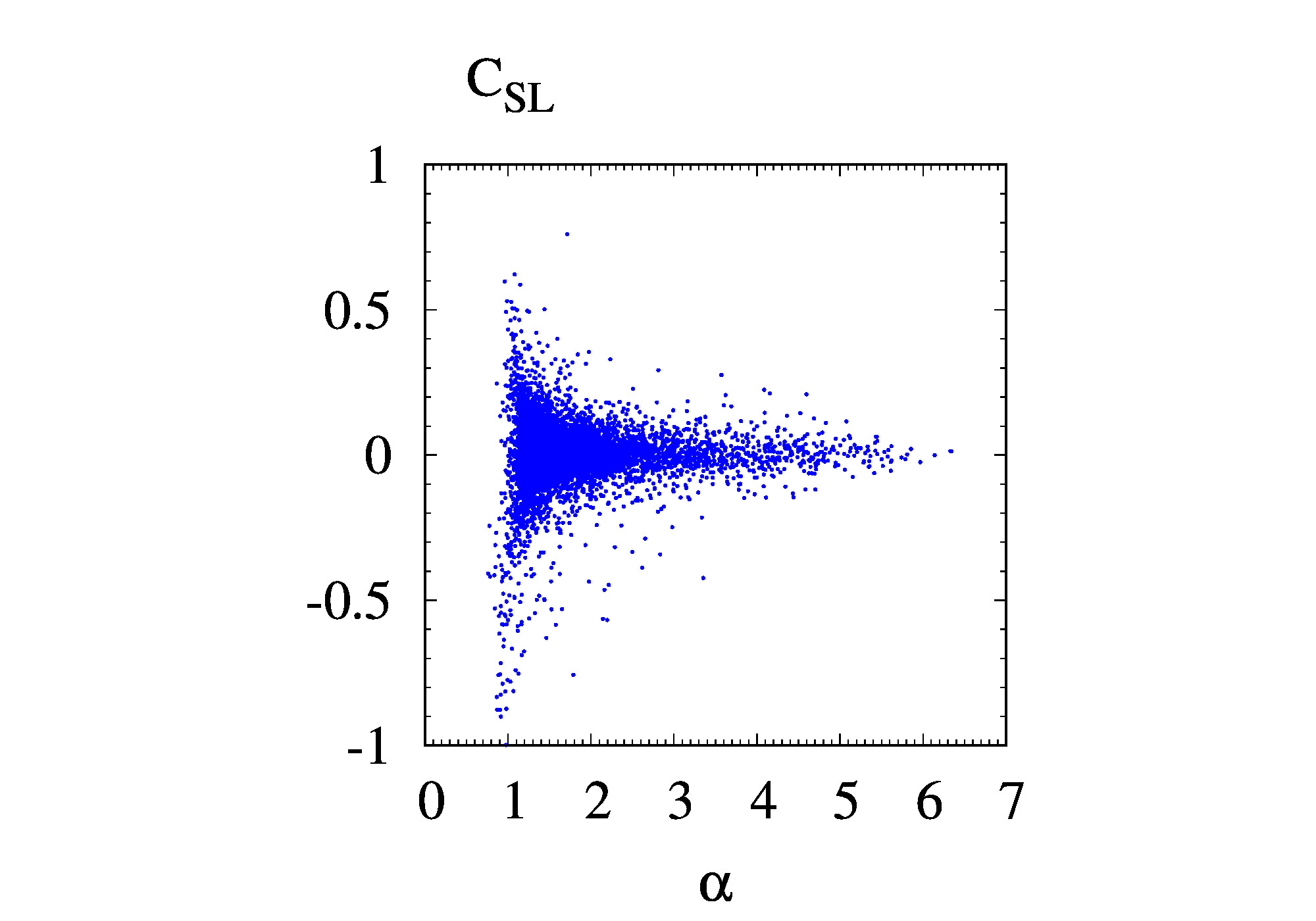}\\
(a) & (b) \\
\hspace{-0.3cm}\includegraphics[scale=0.12]{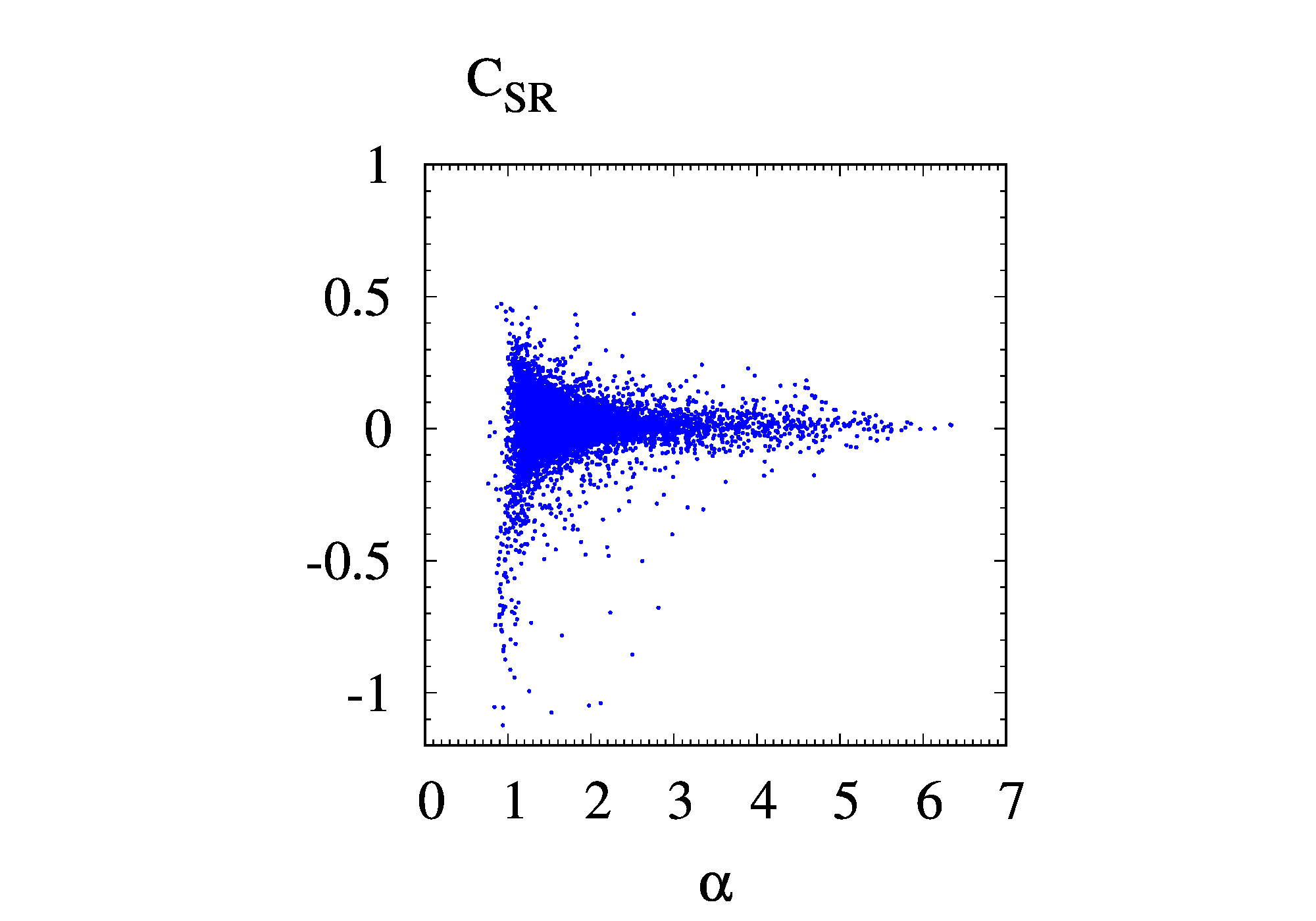}& 
\hspace{-0.3cm}\includegraphics[scale=0.12]{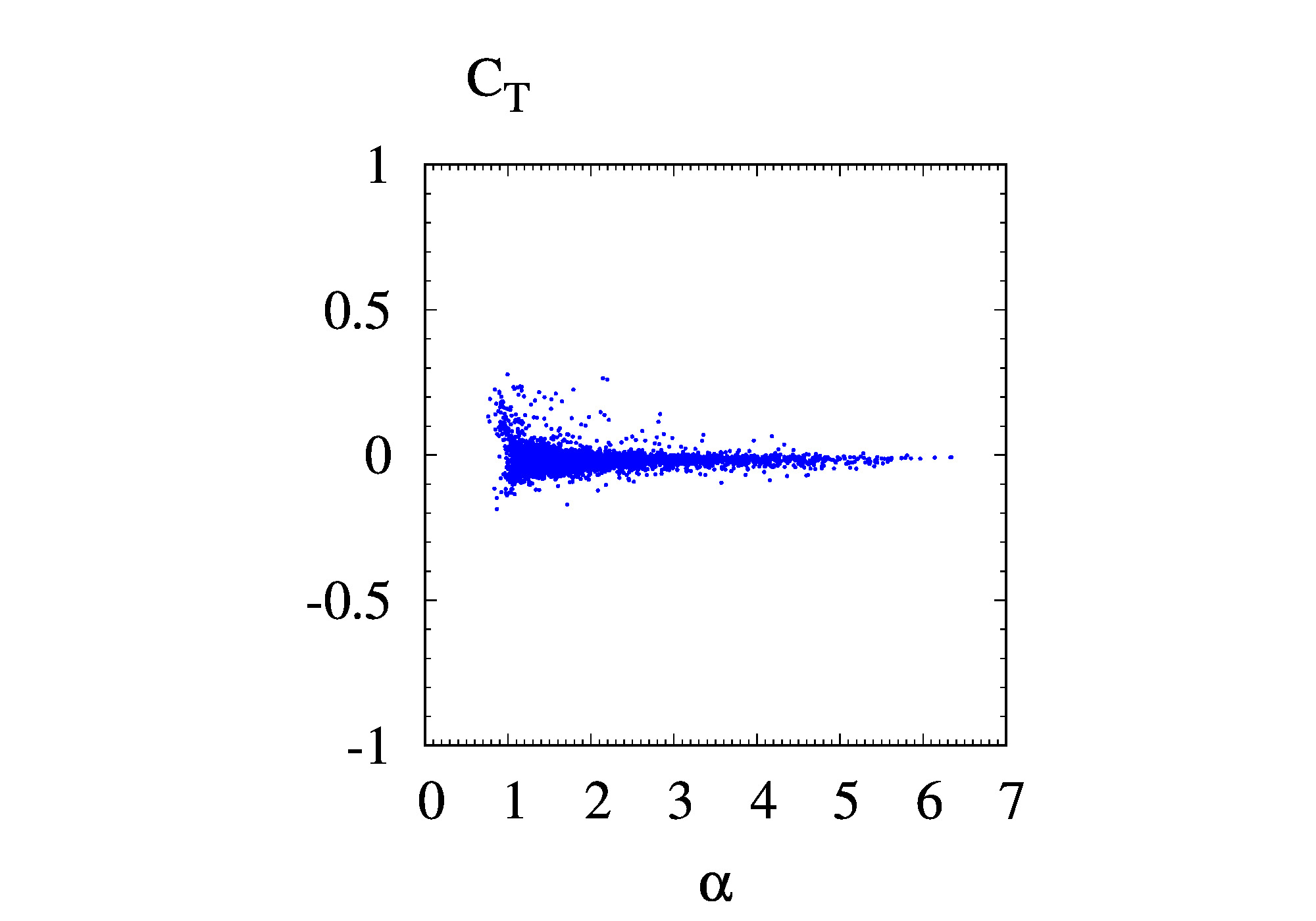} \\
(c) & (d)
\end{tabular}
\caption{\label{alp_C} Allowed regions of (a) $C_{VL}$,
(b) $C_{SL}$, (c) $C_{SR}$, and $C_T$ vs. $\alpha$, respectively,
at the $2\sigma$ level.}
\end{figure}

In Fig.\ \ref{alp_C} we plot allowed regions of $C_j$ with respect to $\alpha$.
%
%----------------- Figure 2 ------------------------------------------------
%----------------------------------------------------------------------------
%
Around $\alpha\approx 1$, $|C_j|$s can cover wider range, and attenuate as $\alpha$ gets larger.
Though our best-fit value of $\alpha$ is around $\sim 2$, $\alpha$ can have $\alpha\gtrsim 6$
as shown in the Figure.
But in our parameter space, $\alpha$ has an upper bound as $\alpha\lesssim 6.4$.
Limiting case of $\alpha\to\infty$ corresponds to the SM.
The fact that $\alpha$ has an upper bound tells that finite NP effects must exist to accommodate
the experimental data; NP effects should not be suppressed so much. 
\par
Note that allowed values of $|C_T|$ are smaller than other $|C_{VL, SL, SR}|$ values. 
The reason is that the coefficients of $C_T$ in Eqs.\ (\ref{RDnum})-(\ref{FLnum}) are
quite larger than those of any other $C_j$s.
Explicitly, we have $-0.186\le C_T \le 0.278$, while 
$-0.568\le C_{VL}\le 0.431$, $-1.000\le C_{SL}\le 0.761$, and
$-1.123\le C_{SR}\le 0.474$.
%
%
%
%----------------- Figure 3  ------------------------------------------------
\begin{figure}
\begin{tabular}{cc}
\hspace{-1cm}\includegraphics[scale=0.12]{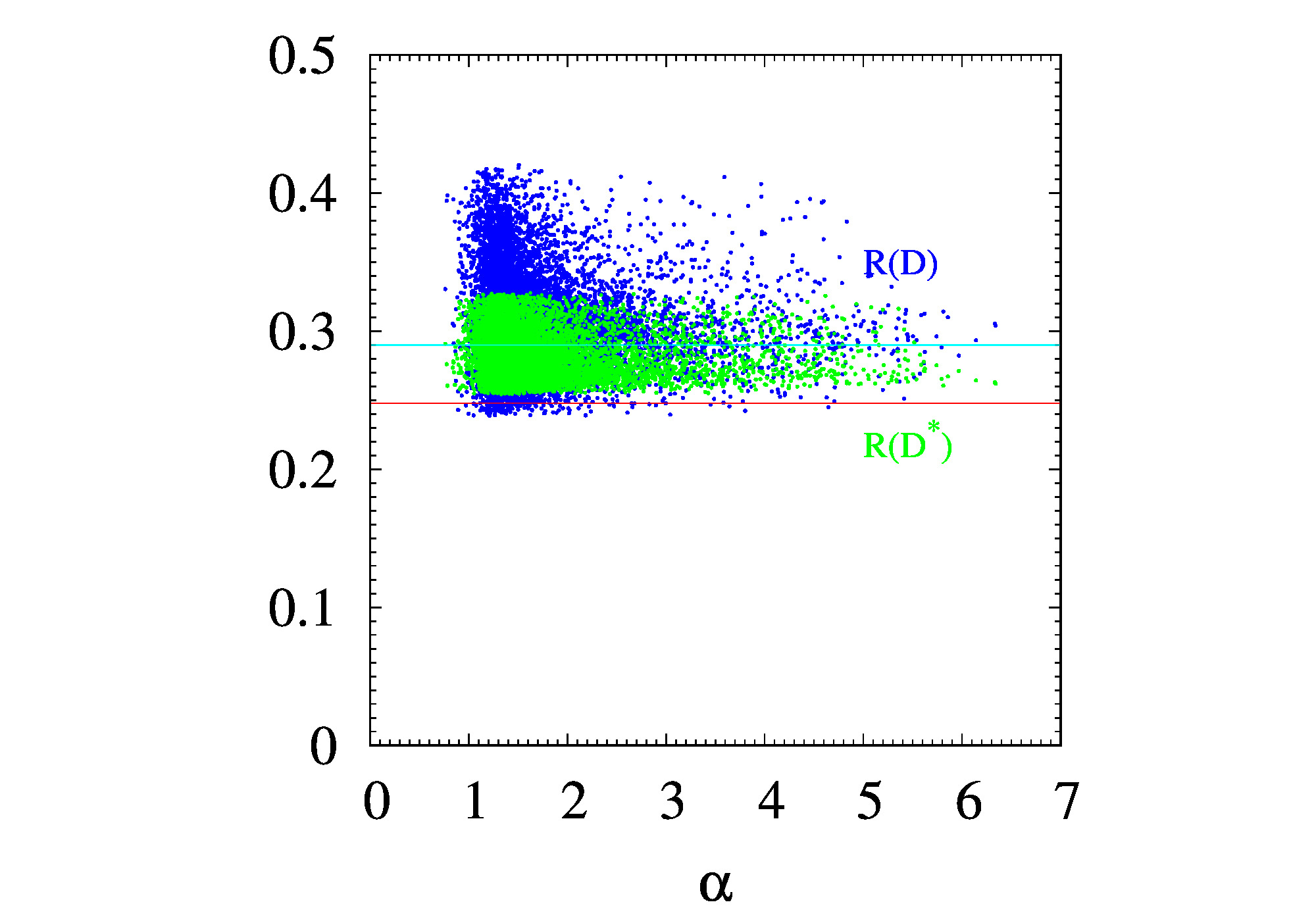}&
\hspace{-1cm}\includegraphics[scale=0.12]{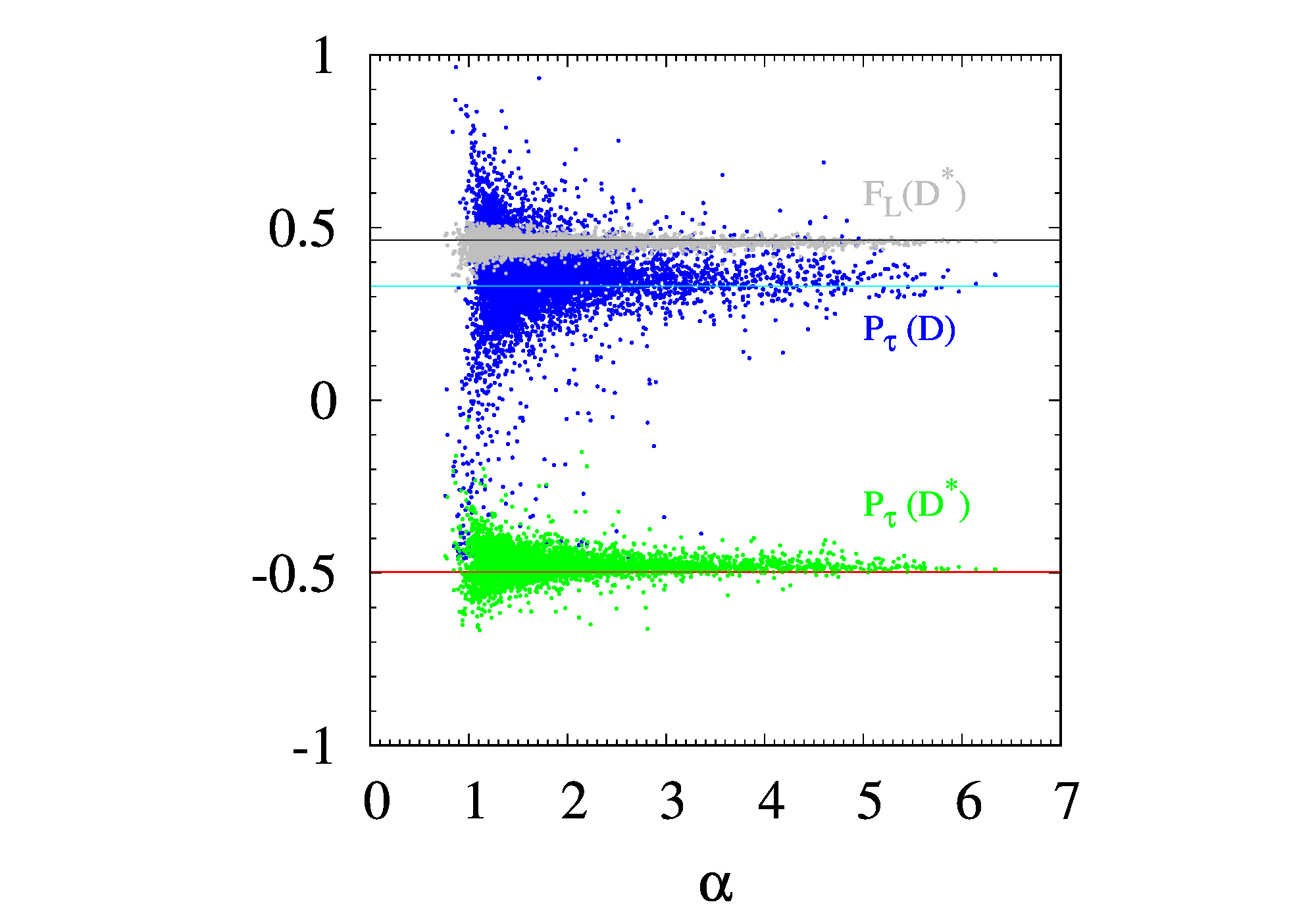}\\
(a) & (b) \\
\hspace{-1cm}\includegraphics[scale=0.12]{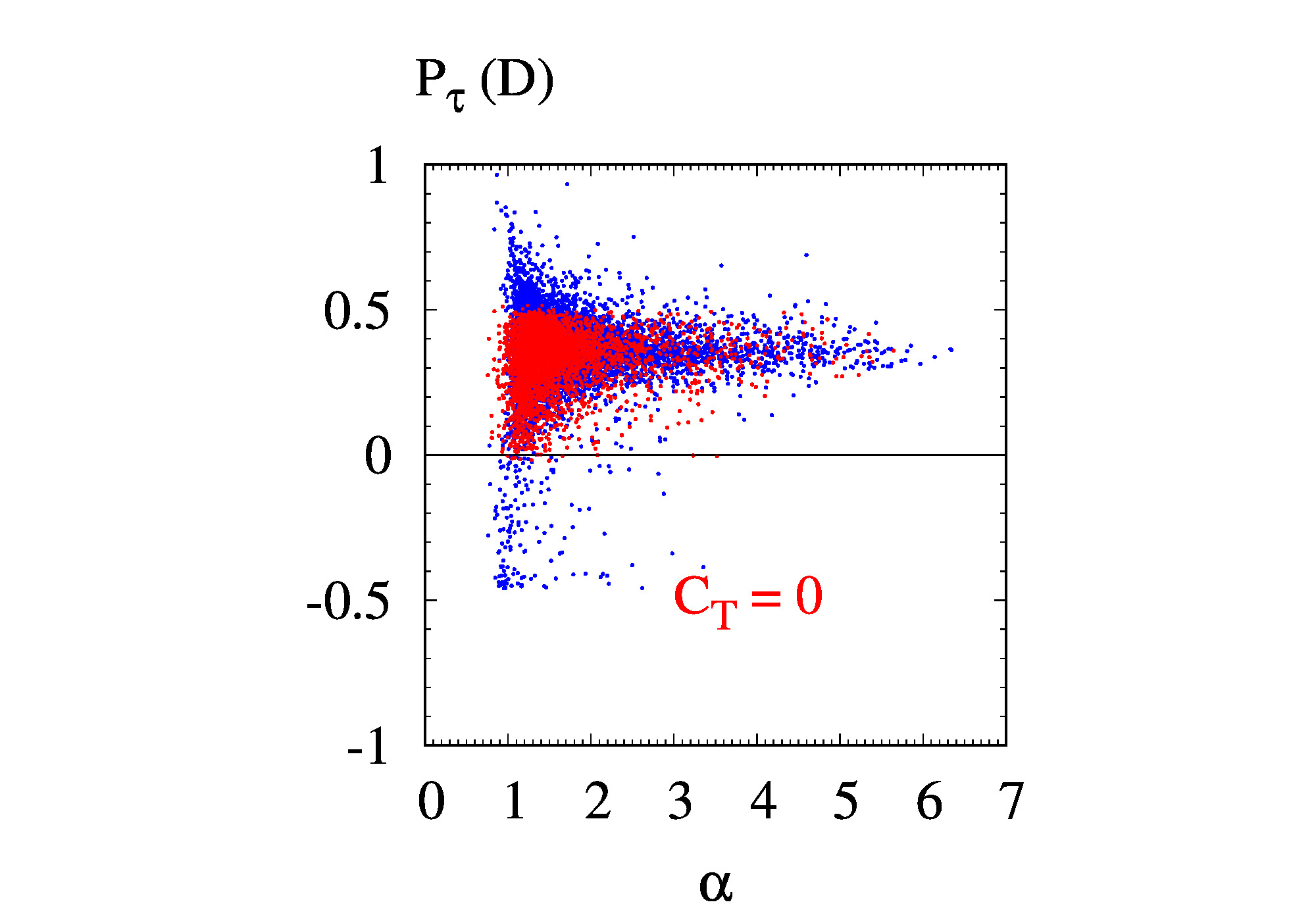}&
\hspace{-1cm}\includegraphics[scale=0.12]{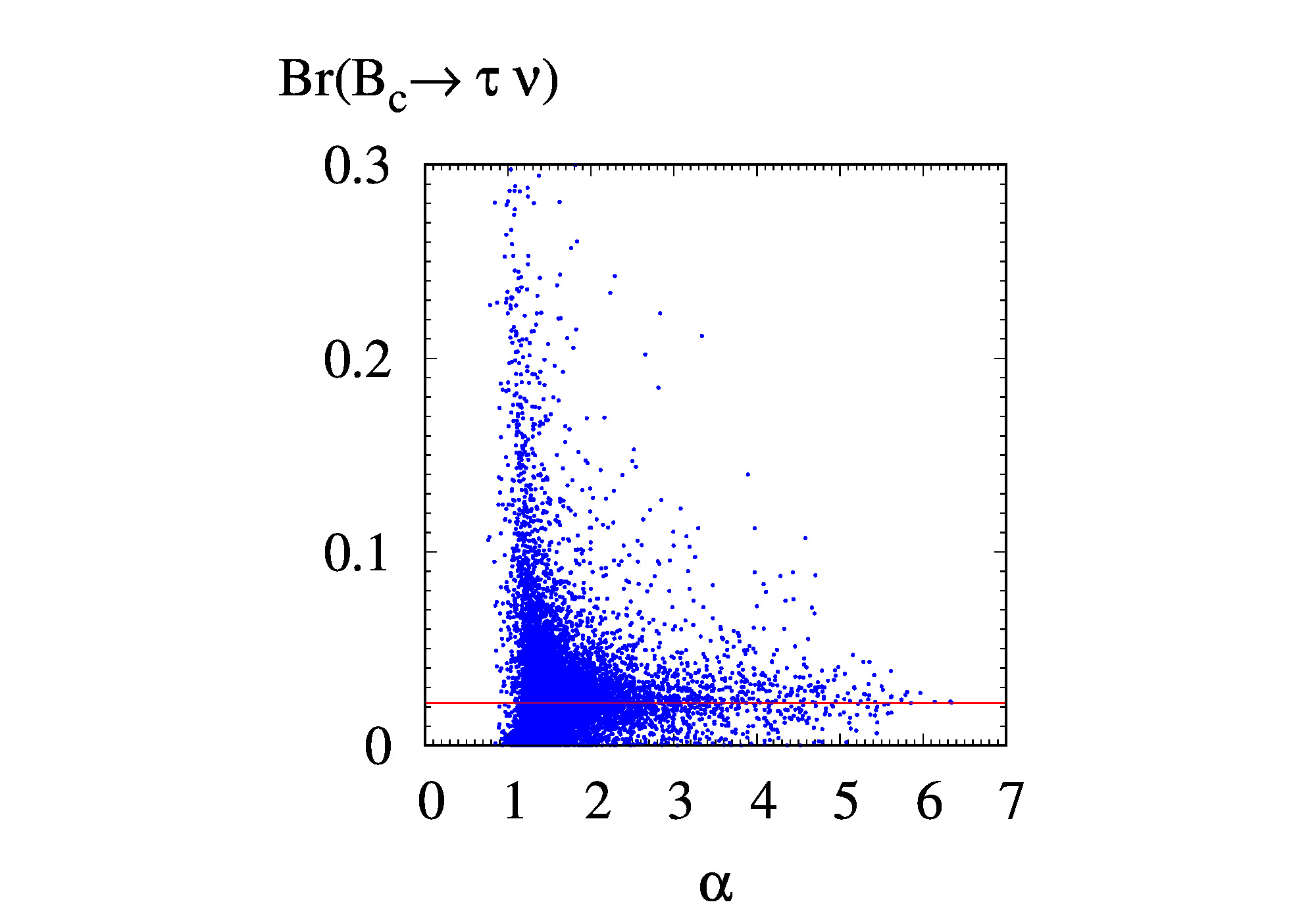} \\
(c) & (d) \\
\end{tabular}
\caption{\label{alp_obs} Allowed regions of 
(a)$R(\Ds)$, (b) $P_\tau(\Ds)$, $F_L(D^*)$,
(c) $P_\tau(D)$ for $C_T\ne 0$ (blue) and $C_T=0$ (red),
and (d) $\Br(B_c\to\tau\nu)$ vs. $\alpha$, respectively, 
at the $2\sigma$ level.
In (a) cyan(red) line stands for the SM prediction of $R(D)(R(D^*))$;
in (b) grey, cyan, and red lines stand for the SM prediction of $F_L(D^*), P_\tau(D)$, 
and $P_\tau(D^*)$, respectively;
in (d) red line is the SM prediction of the branching ratio.
}
\end{figure}
%----------------------------------------------------------------------------
%
\par
Figure \ref{alp_obs} shows the allowed regions of various observables,
$R(\Ds), P_\tau(\Ds), F_L(D^*)$, and $\Br(B_c\to\tau\nu)$, with respect to $\alpha$.
As in Fig.\ \ref{alp_C}, these observables span wider range for $\alpha\sim 1$.
As can be seen in Fig.\ \ref{alp_obs} (a) and (b), 
$R(D), P_\tau(\Ds), F_L(D^*)$, and $\Br(B_c\to\tau\nu)$ tends to converge to the SM predictions 
as $\alpha$ gets larger in our fittings.
But $R(D^*)$ values are still deviated from the SM prediction for large $\alpha$.
It reflects the fact that $R(D^*)$ data are farther from the SM value.
In Fig. \ref{alp_obs} (b) 
one can find that our fits have much more plots in positive $P_\tau(D)$ than negative ones,
but negative $P_\tau(D)$ is also possible.
The SM prediction for $P_\tau(D)$ is positive definite as given in Eq.\ (\ref{PDSM}),
so the observation of negative $P_\tau(D)$ would strongly suggest the existence of NP.
This is quite different from the unparticle scenario results.
In Ref. \cite{JPL2012} it was shown that unparticle contributes to produce positive $P_\tau(D)$
(see Fig.\ 2(c) therein).
\par
When turning on and off various combinations of $C_{VL, SL, SR, T}$, 
we find that only two cases where 
(A) $C_{VL,SR}\ne 0$ and (B) $C_{VL,SL,T}\ne 0$ 
allow negative $P_\tau(D)$.
In case of (A) if $C_T=0$ then it is somewhat marginal such that
almost all the allowed points of $P_\tau(D)$ are positive and a few points are slightly negative;
see Fig.\ \ref{alp_obs} (c).
If $P_\tau(D)$ turned out to be negative, then it could restrict the shape of NP very strongly.
Especially, if $P_\tau(D)$ is negatively large, then $C_T$ must be nonzero. 
\par
Figure \ref{alp_obs} (c) shows that most of $\Br(B_c\to\tau\nu)$ values are safely less than $\sim 0.1$. 
Note that our constraint is $\Br(B_c\to\tau\nu)<0.3$.
Since the SM value of the branching ratio (=$0.022$) is quite small compared to the upper limit,
the constraint on the branching ratio is not so strong.
%
%----------------- Figure 4 : fixed alpha=2 ------------------------------------------------
\begin{figure}
\begin{tabular}{cc}
\hspace{-1cm}\includegraphics[scale=0.12]{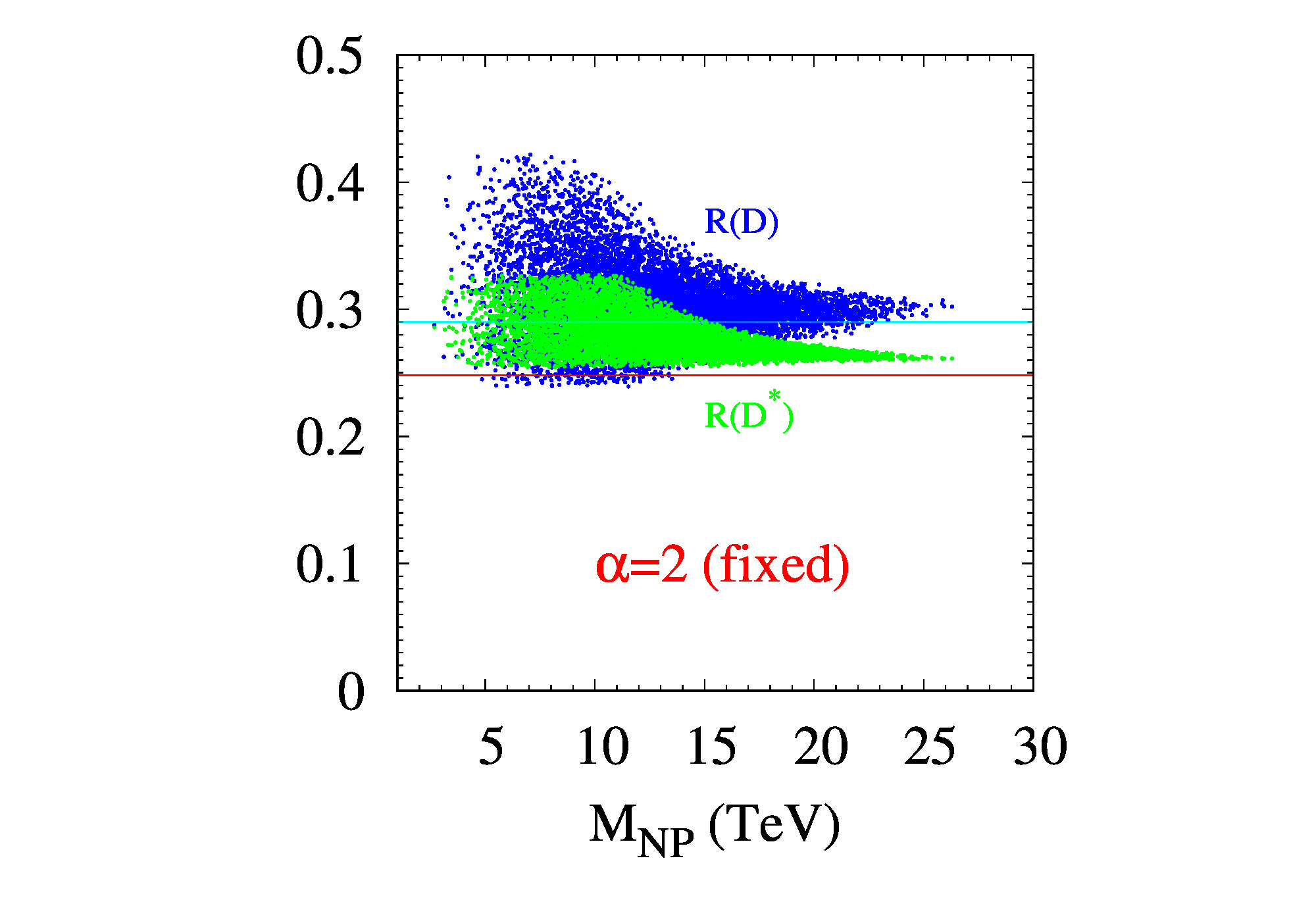}&
\hspace{-1cm}\includegraphics[scale=0.12]{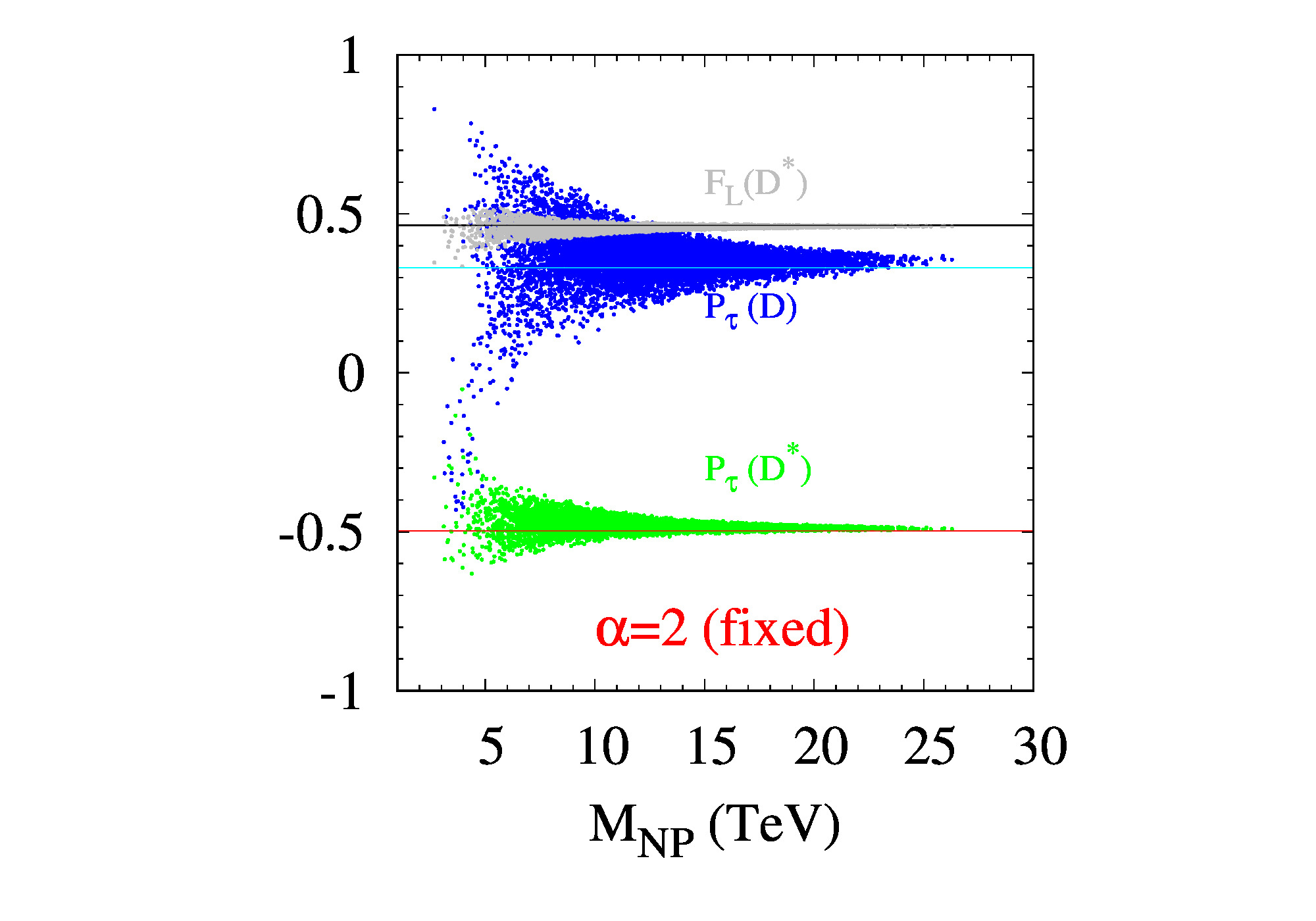}\\
(a) & (b) \\
\hspace{-1cm}\includegraphics[scale=0.12]{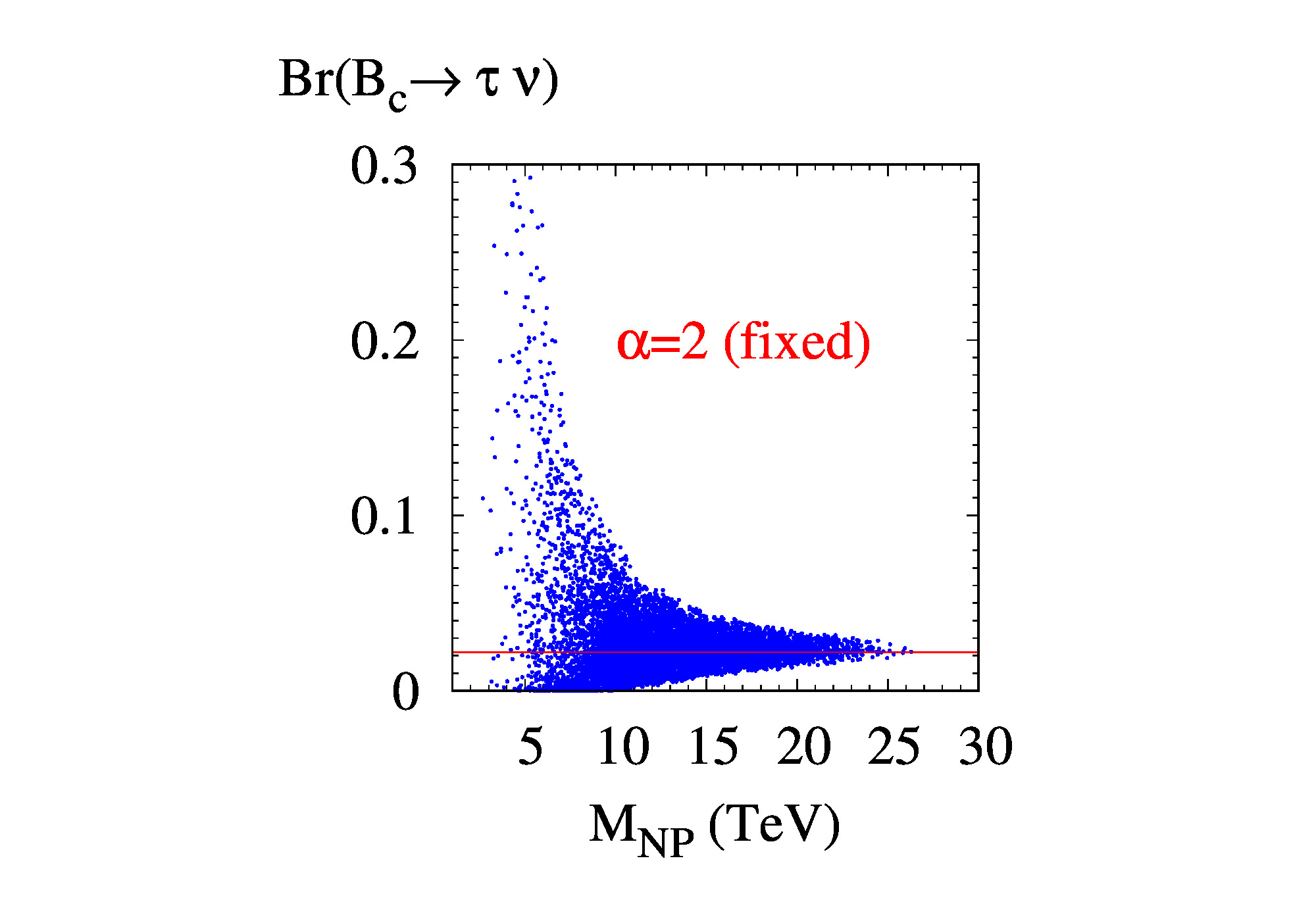}& \\
(c) & \\
\end{tabular}
\caption{\label{m00_obs} Allowed regions of 
(a)$R(\Ds)$, (b) $P_\tau(\Ds)$, $F_L(D^*)$,
and (c) $\Br(B_c\to\tau\nu)$ vs. $M_\NP$, respectively, 
at the $2\sigma$ level for fixed $\alpha=2$.
In (a) cyan(red) line stands for the SM prediction of $R(D)(R(D^*))$;
in (b) grey, cyan, and red lines stand for the SM prediction of $F_L(D^*), P_\tau(D)$, 
and $P_\tau(D^*)$, respectively;
in (c) red line is the SM prediction of the branching ratio.
}
\end{figure}
%------------------------------------------------------------------------------------------------------------------------
%
\par
In Fig.\ \ref{m00_obs} various observables are plotted with respect to $M_\NP$ for fixed $\alpha=2$.
All the observables shown tend to attenuate close to the SM values as $M_\NP$ gets larger.
One can clearly see again that $M_\NP\lesssim 27~\TeV$ for $\alpha=2$.
For larger values of $M_\NP$, NP effects get smaller and could not be compatible with the experimental data.
The limiting case of $M_\NP\to\infty$ corresponds to the SM.
Thus the upper bound of $M_\NP$ (for $\alpha=2$) indicates the existence of NP.
The figures show that NP should exist to explain the experimental data in case of $\alpha=2$.
For smaller $\alpha\lesssim 1$, $M_\NP$ can be much larger (see Fig.\ \ref{alp_M})
where NP effects would not be suppressed so much.
As mentioned before, negative $P_\tau(D)$ would indicate a strong evidence of NP.
If NP is mediated by ordinary particles where $\alpha=2$ and $P_\tau(D) <0$,
then $M_\NP$ could not be larger than $\simeq 6~\TeV$ as can be seen in Fig.\ \ref{m00_obs} (b).
%
%
%
%----------------- Figure 5 : Wilson coefficients   ------------------------------------------------
\begin{figure}
\begin{tabular}{cc}
\hspace{-1cm}\includegraphics[scale=0.12]{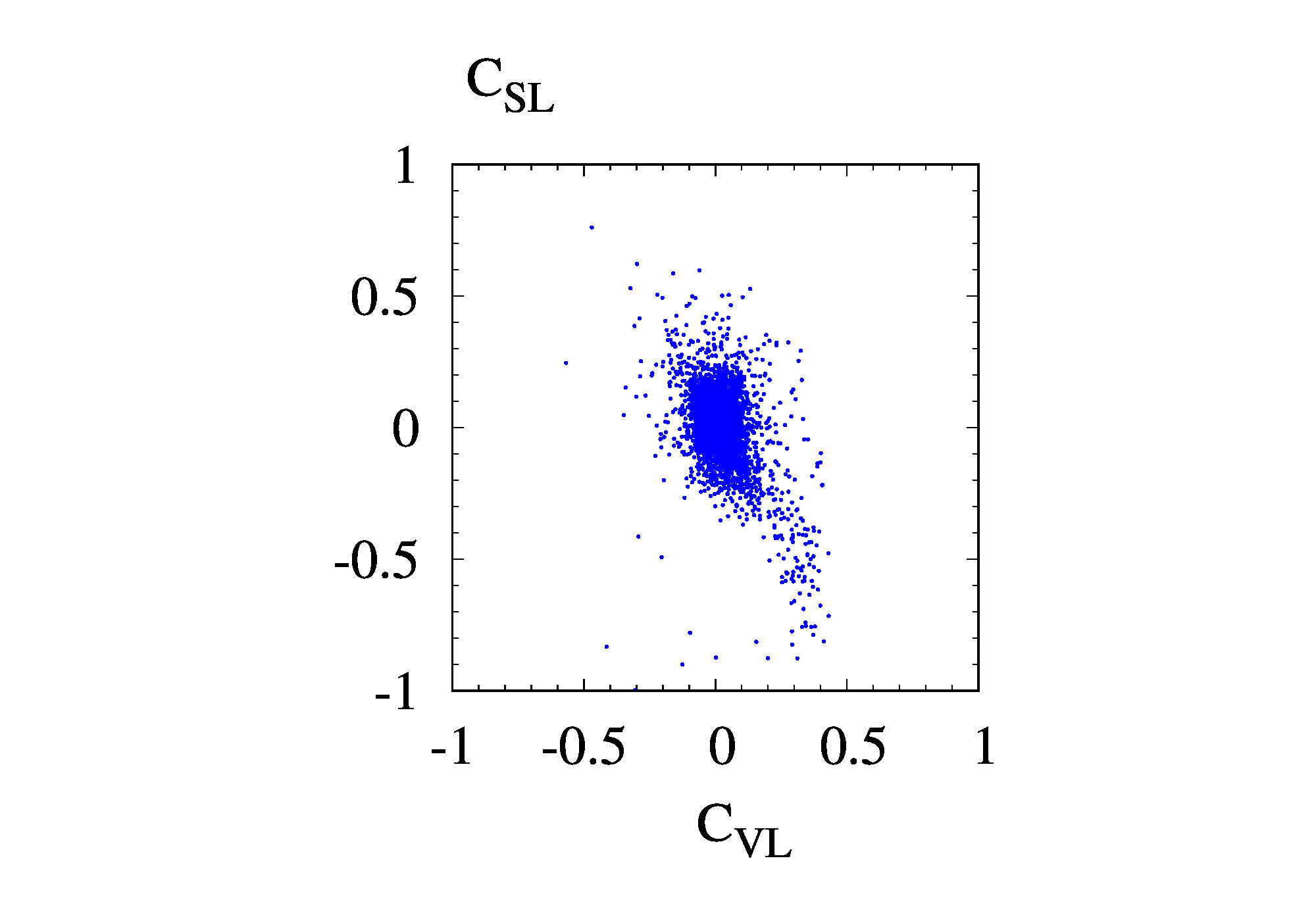}&
\hspace{-1cm}\includegraphics[scale=0.12]{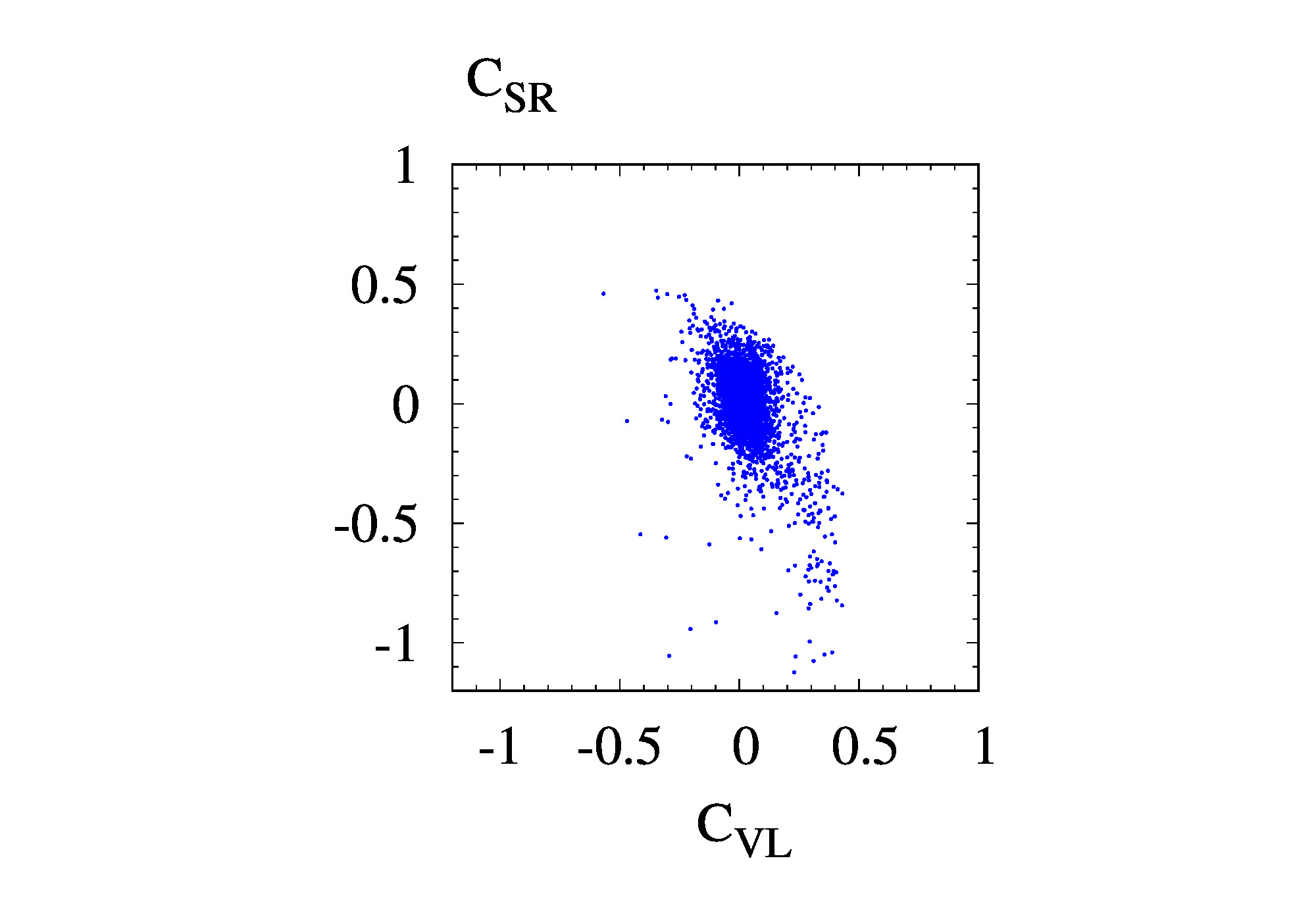}\\
(a) & (b) \\
\hspace{-1cm}\includegraphics[scale=0.12]{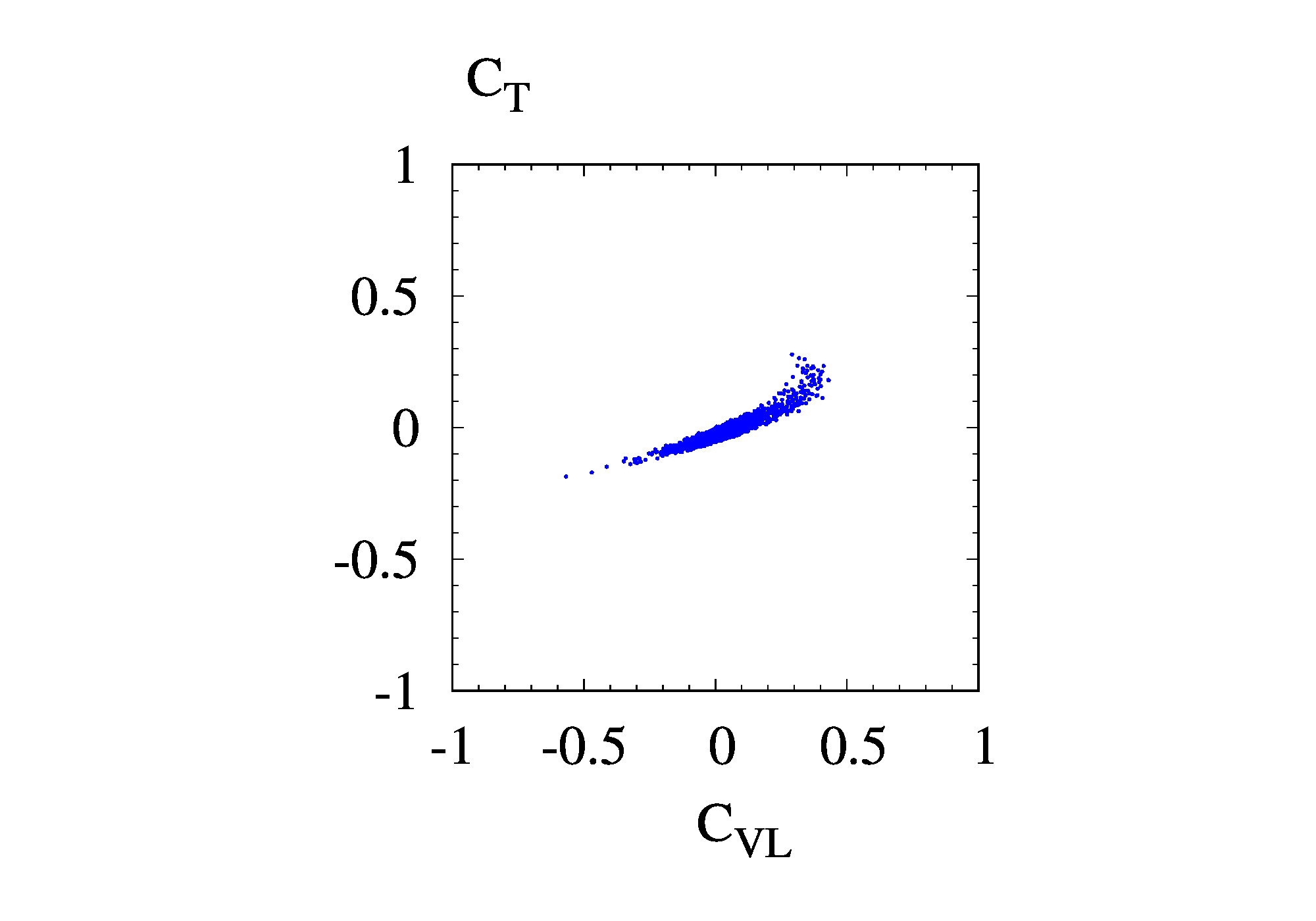}&
\hspace{-1cm}\includegraphics[scale=0.12]{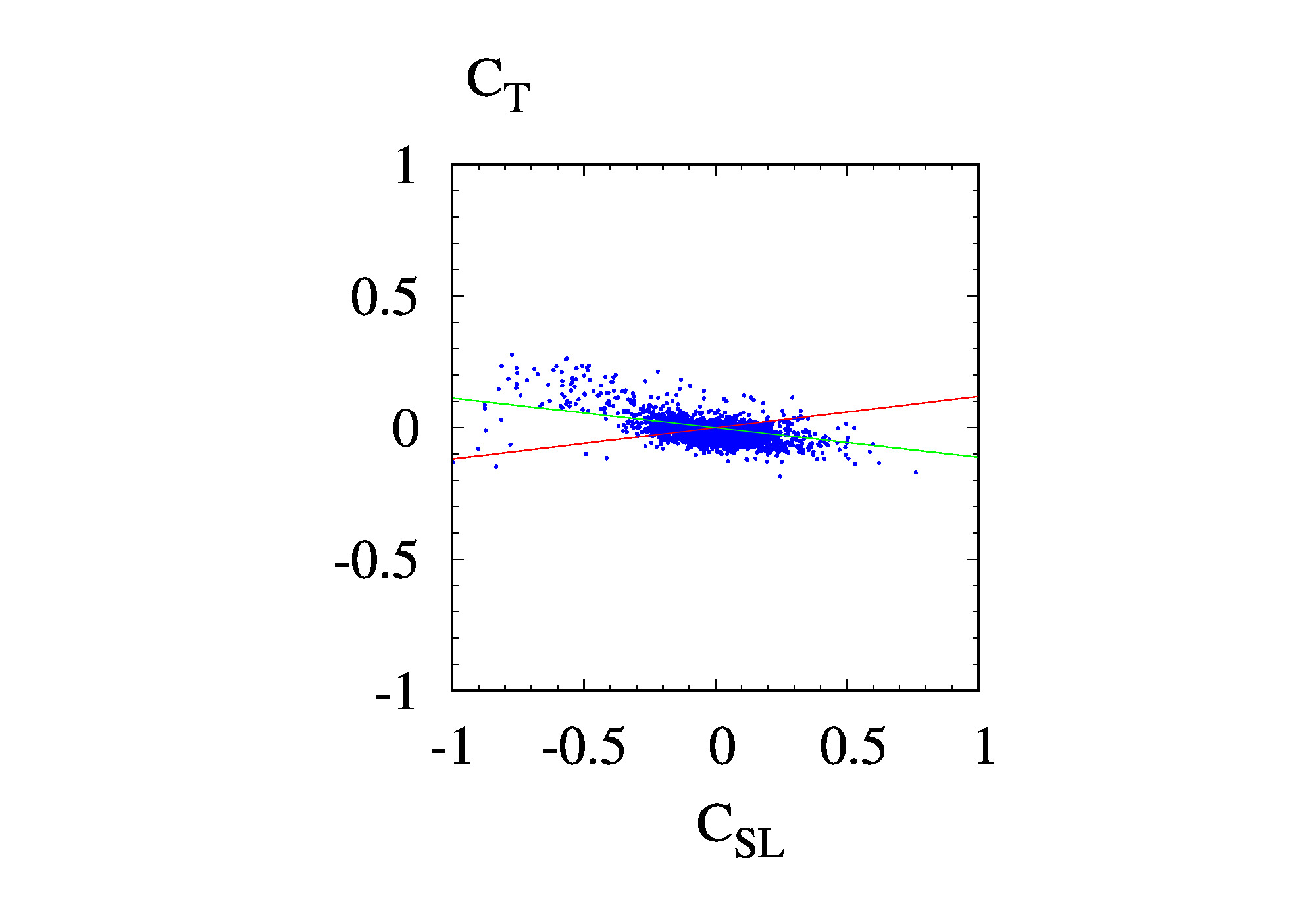}\\
(c) & (d) \\
\hspace{-0.3cm}\includegraphics[scale=0.12]{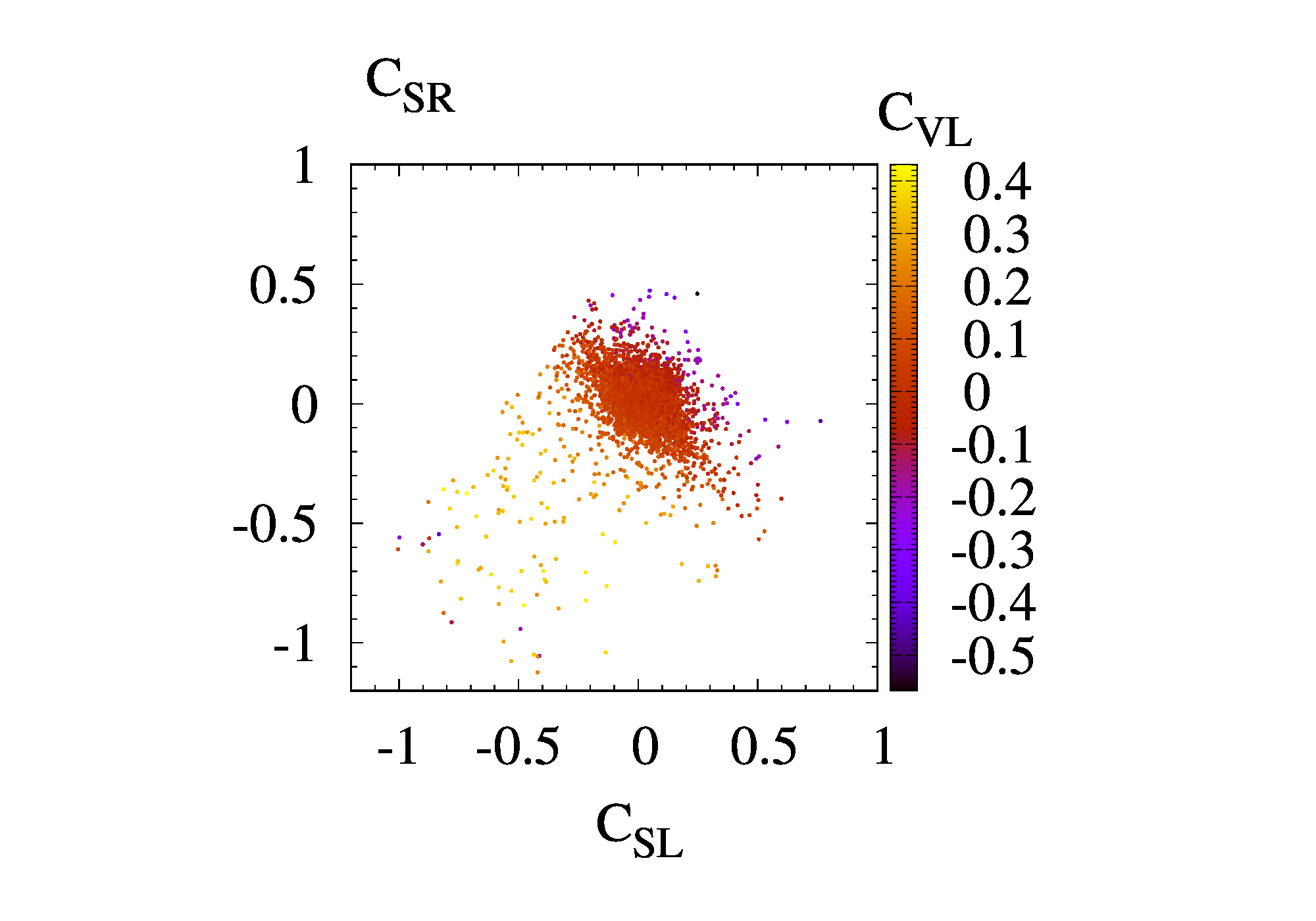}& \\
(e)
\end{tabular}
\caption{\label{WC} Allowed regions of 
(a) $C_{SL}$ vs. $C_{VL}$, (b) $C_{SR}$ vs. $C_{VL}$, 
(a) $C_{T}$ vs. $C_{VL}$, (b) $C_{T}$ vs. $C_{SL}$,
and (e) $C_{SR}$ vs. $C_{SL}^\tau$ with respect to $C_{VL}^\tau$
at the $2\sigma$ level.
In (d), the red(green) line corresponds to $C_{SL}=+8.4 C_T (-8.9 C_T)$
at $\mu=\mu_b$.}
\end{figure}
%----------------------------------------------------------------------------------------------------
%
\par
In Fig.\ \ref{WC} we plot the Wilson coefficients $C_{VL, SL, SR, T}$.
Among 4 Wilson coefficients $C_{VL, SL, SR, T}$ in out analysis
two or three of them can be zero, but not all of them.
It means that the SM where $C_j=0$ is not included in our allowed region at the $2\sigma$.
In Fig\ \ref{WC} (d), straight lines represent 
$C_{SL}(\mu_b)=+8.4 C_T(\mu_b)$ and $C_{SL}(\mu_b)=-8.9 C_T(\mu_b)$,
which are inspired by the leptoquark model \cite{Iguro2210,Capdevila2309}.
As seen in Fig.\ \ref{WC} (d), our result prefers $C_{SL}= -8.9 C_T$.
%
%
%----------------- Figure 6 : fixed alpha=2 ------------------------------------------------
\begin{figure}
\begin{tabular}{cc}
\hspace{-1cm}\includegraphics[scale=0.12]{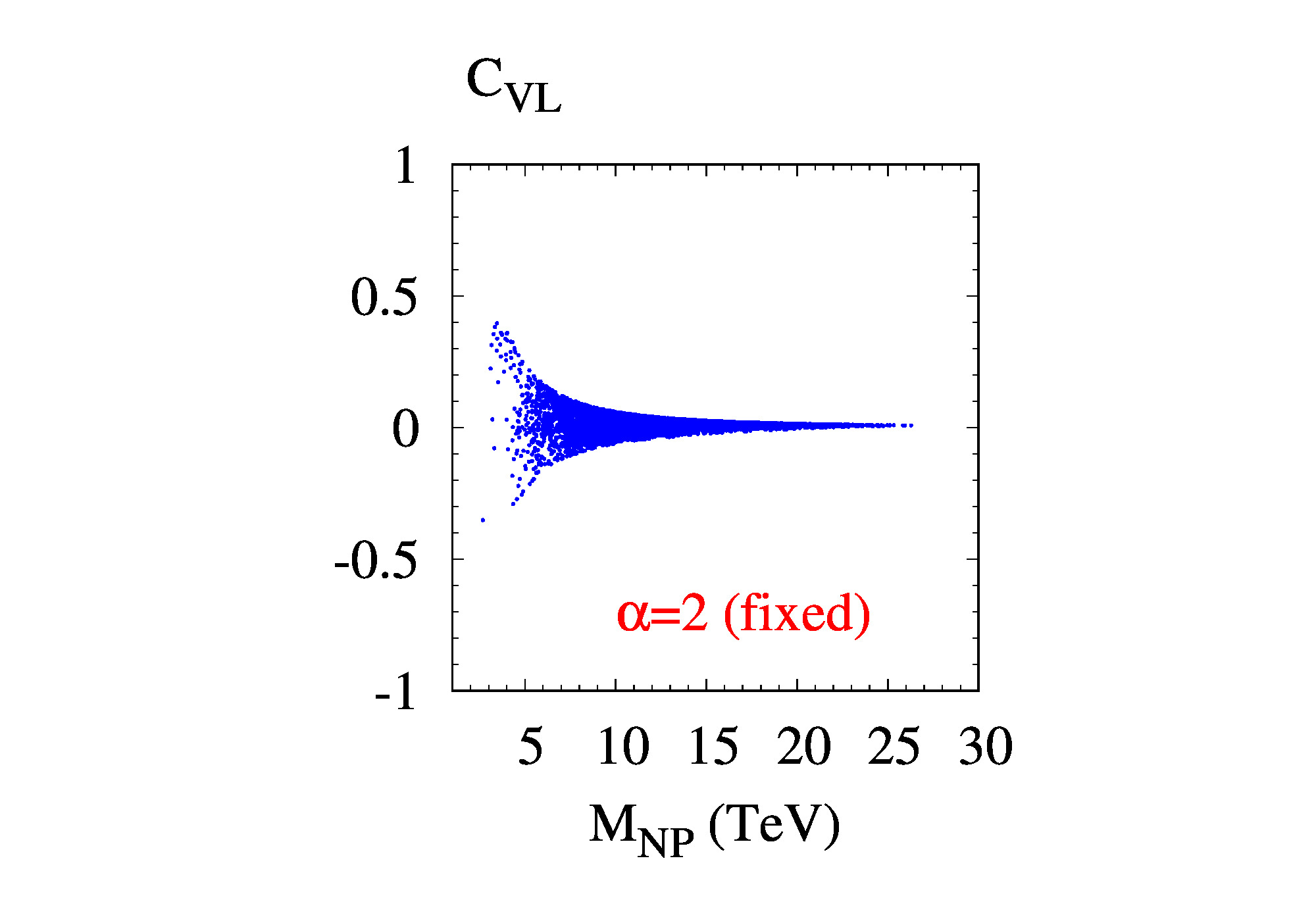}&
\hspace{-1cm}\includegraphics[scale=0.12]{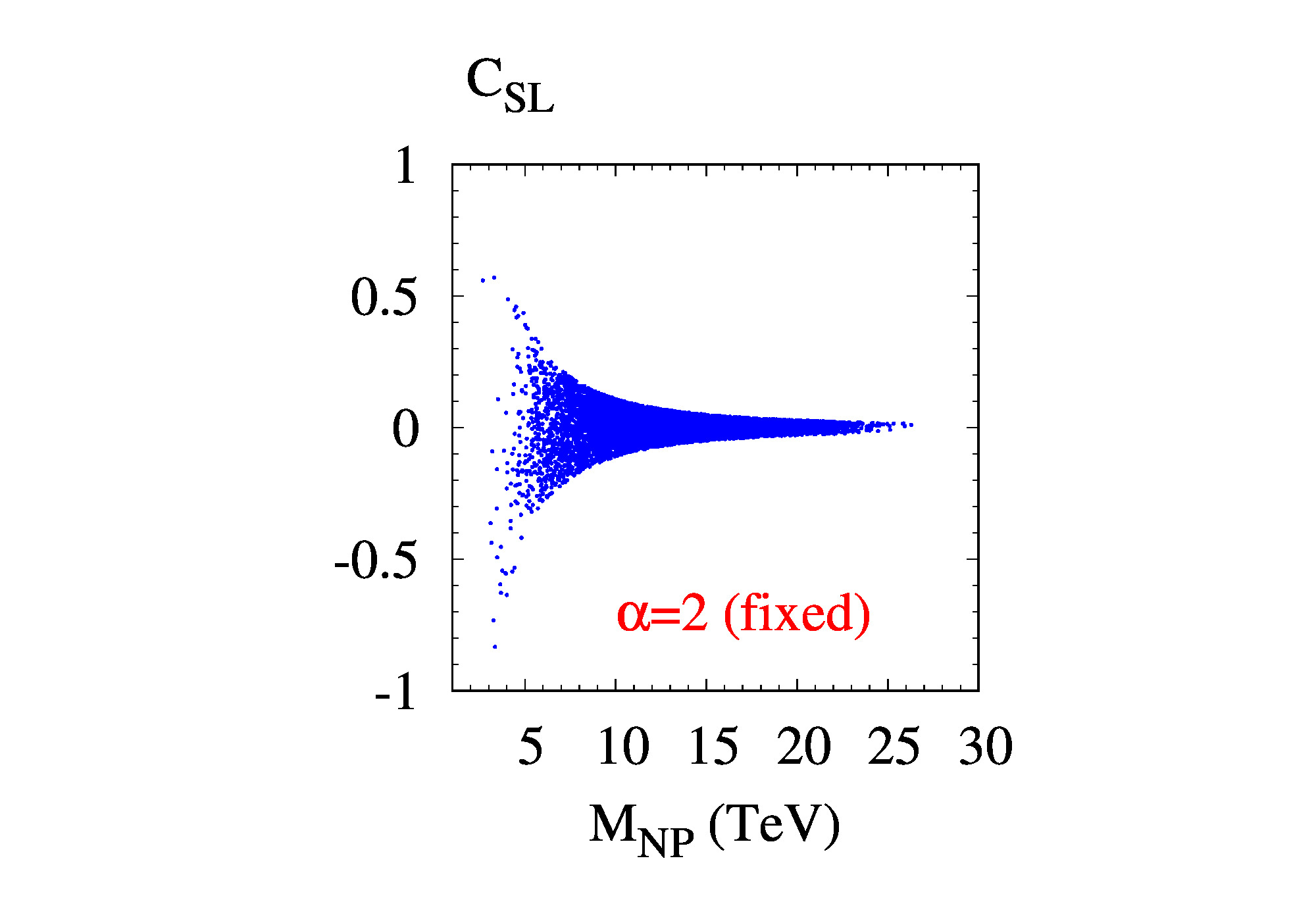} \\
(a) & (b) \\
\hspace{-0.3cm}\includegraphics[scale=0.12]{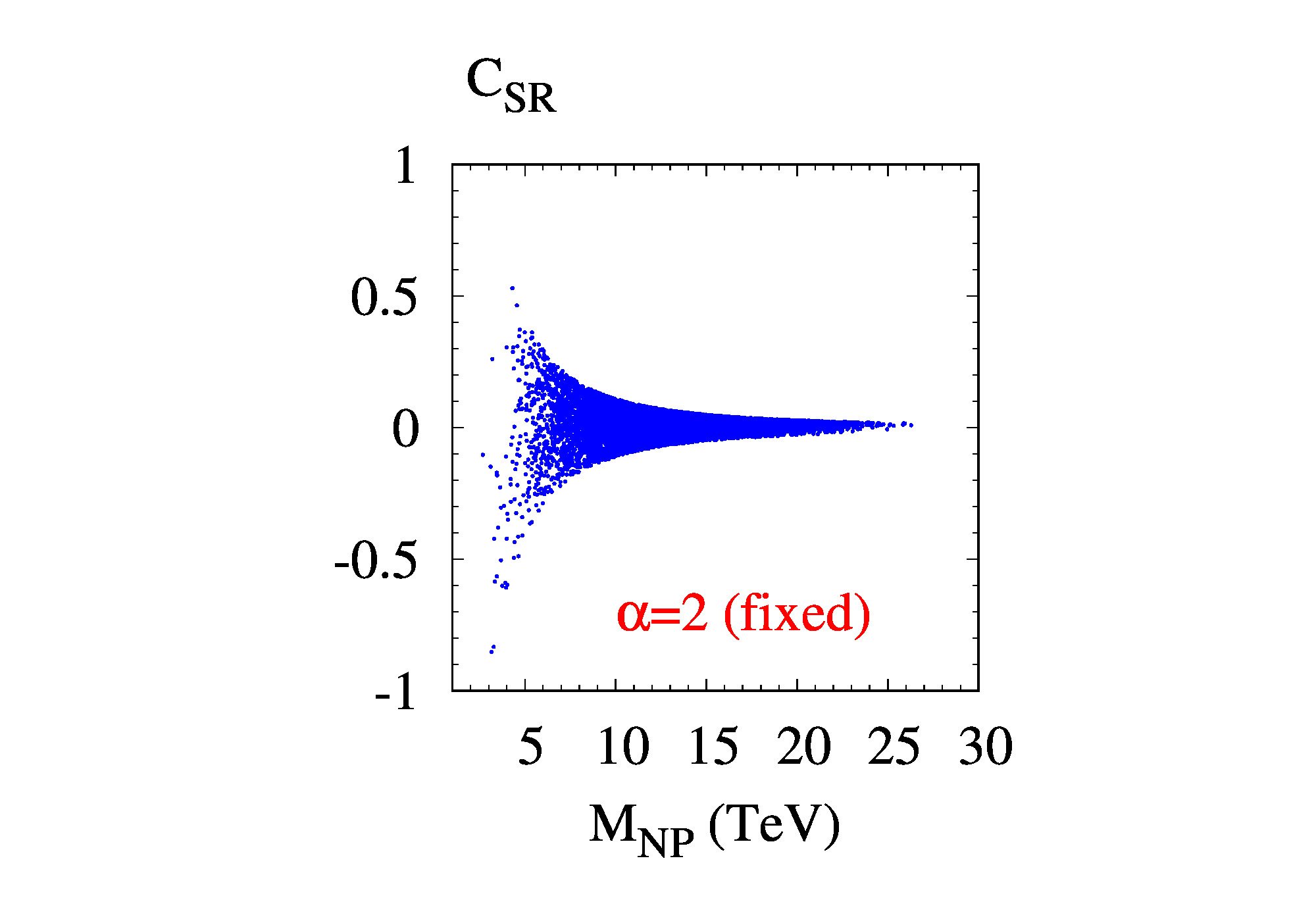}& 
\hspace{-0.3cm}\includegraphics[scale=0.12]{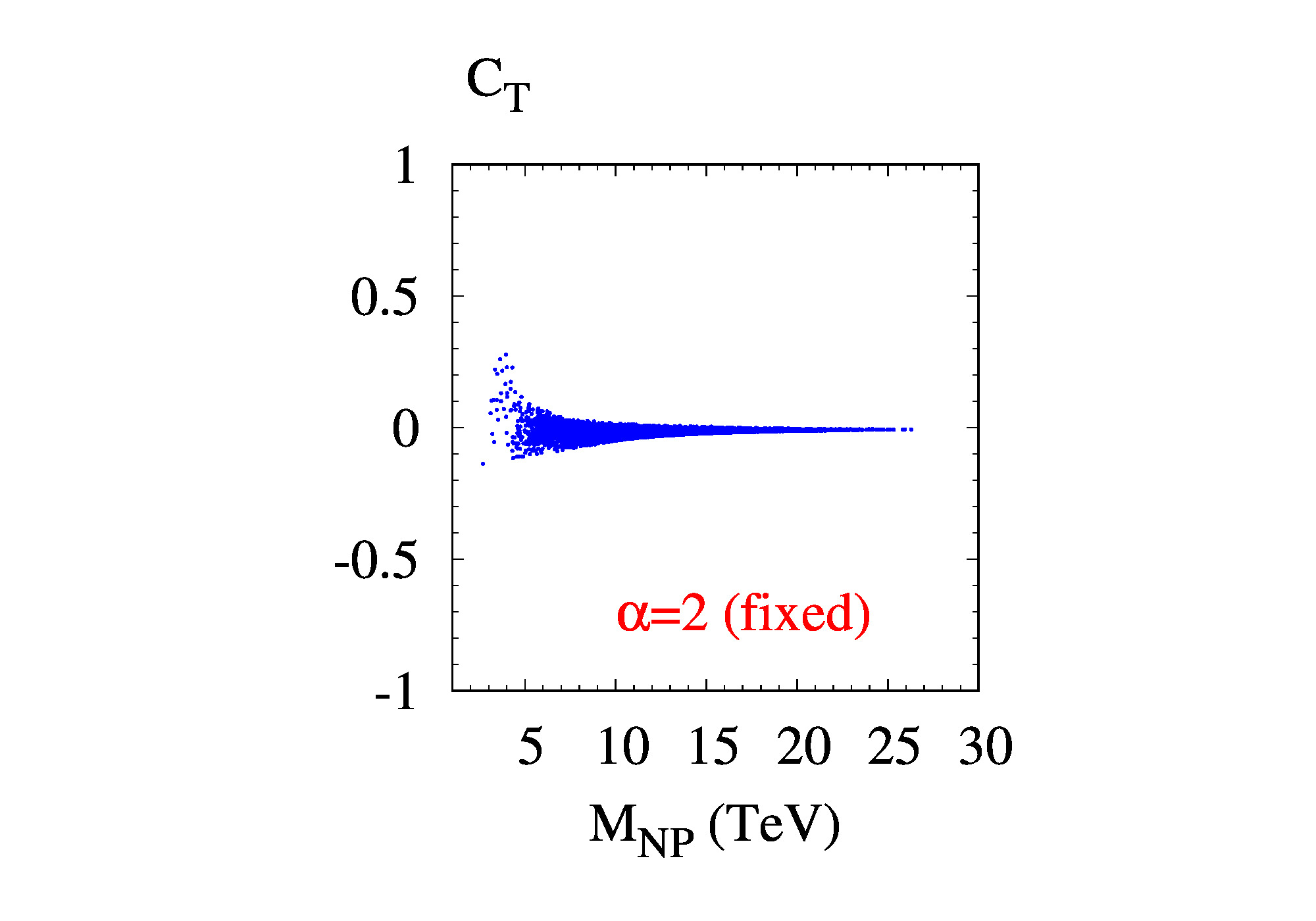} \\
(c) & (d)
\end{tabular}
\caption{\label{m00_C} Allowed regions of 
(a) $C_{VL}$, (b) $C_{SL}$,
(c) $C_{SR}$, and (d) $C_T$ vs. $M_\NP$, respectively,
at the $2\sigma$ level for fixed $\alpha=2$.}
\end{figure}
%------------------------------------------------------------------------------------------------------------------
%
\par
Figure \ref{m00_C} plots the Wilson coefficients vs. $M_\NP$ for the case of fixed $\alpha =2$.
The plots show a typical new-"particle" contribution of $C_j\sim 1/M_\NP^2$.
The $M_\NP$ dependences of $C_{j\ne VL}$ involve the RG running effects.
Comparing Fig.\ \ref{m00_C} (a) with (b) or (c), we find that the RG effects are not dominant.
Again one can see that for fixed $\alpha=2$, $M_\NP$ can be as large as $\sim 27 ~\TeV$.

%
%
%
%
%----------------- Figure 7: Observables  ------------------------------------------------
\begin{figure}
\begin{tabular}{cc}
\hspace{-1cm}\includegraphics[scale=0.12]{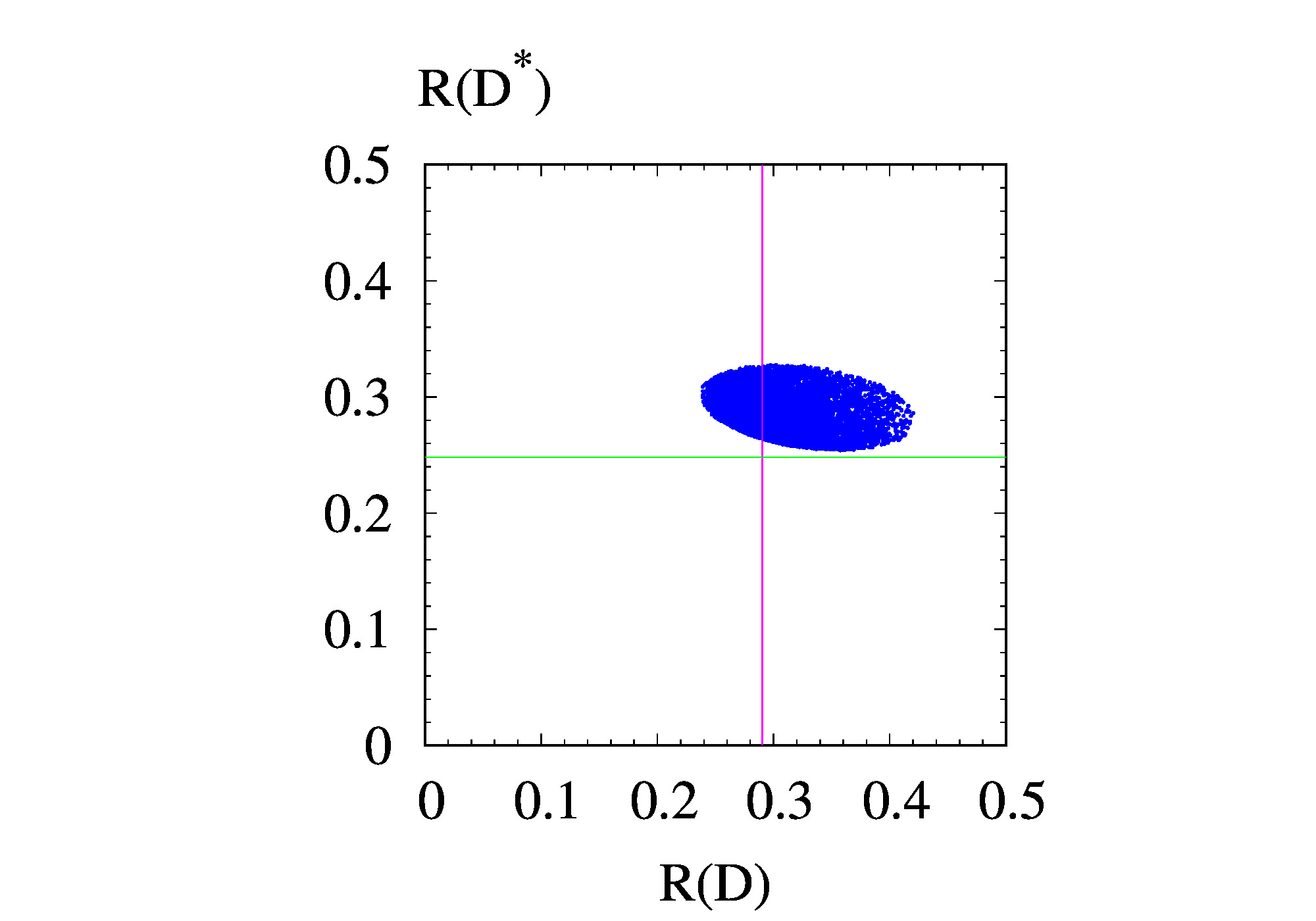} &
\hspace{-1cm}\includegraphics[scale=0.12]{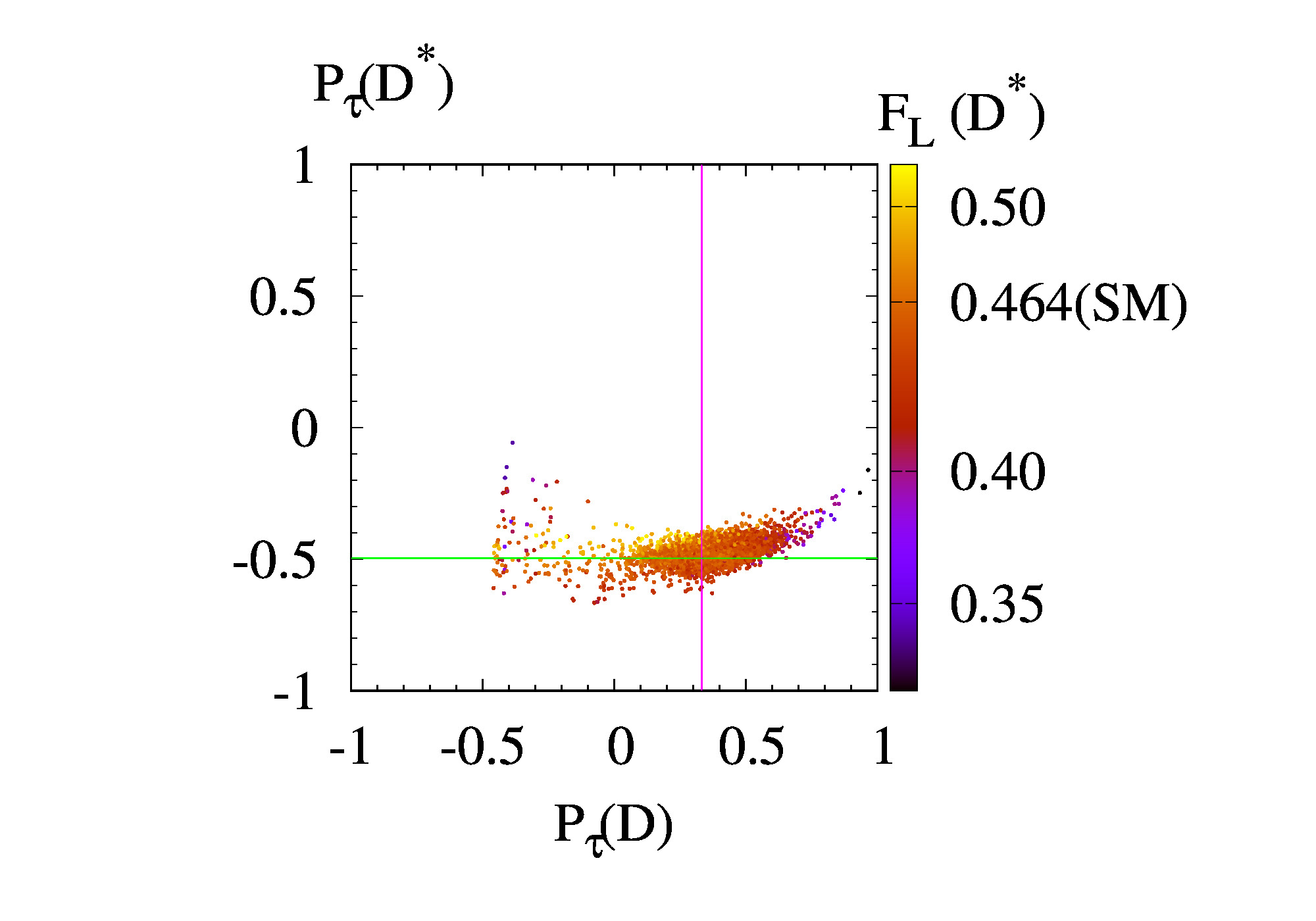} \\
(a) & (b) \\
\hspace{-1cm}\includegraphics[scale=0.12]{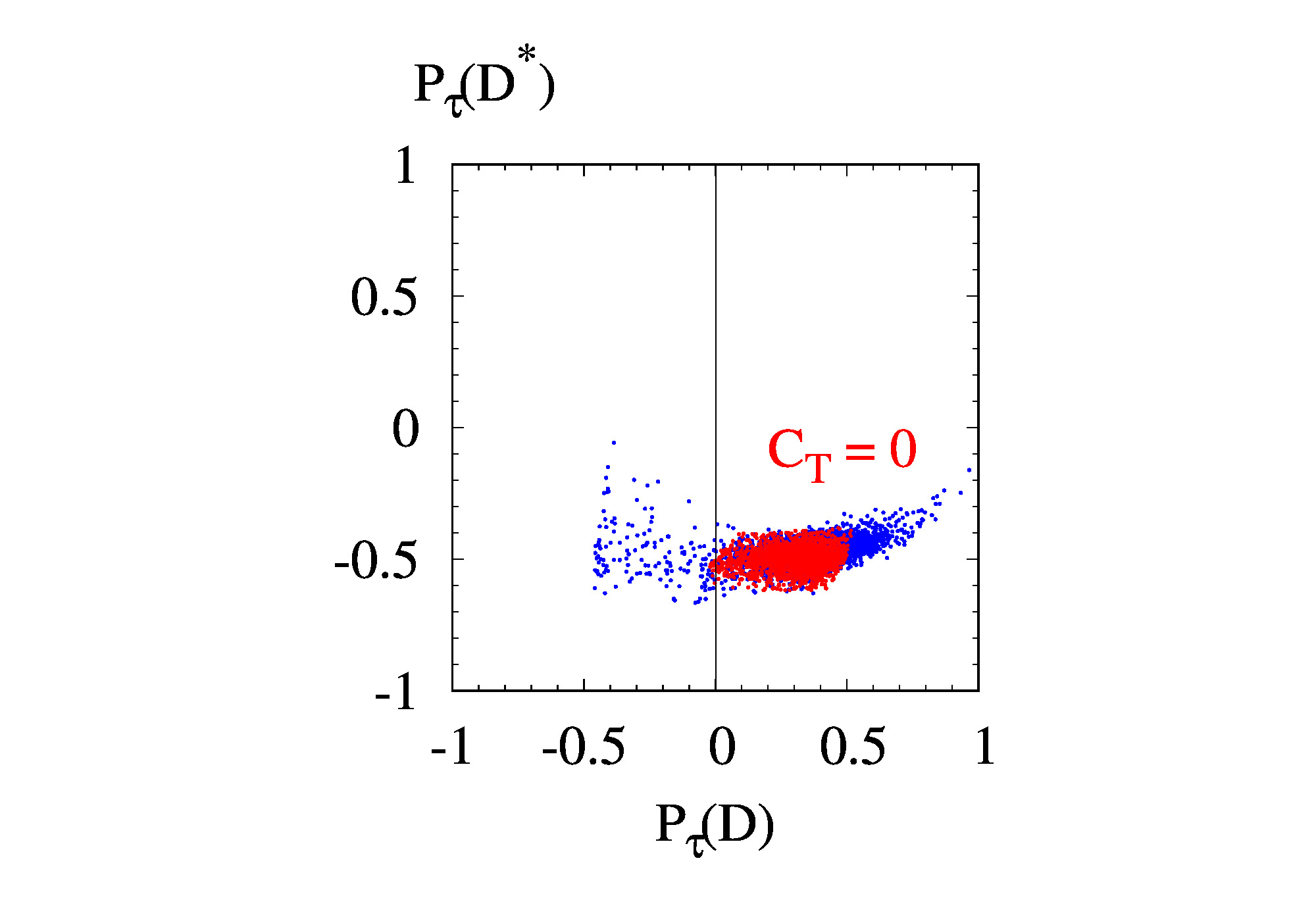} &
\hspace{-1cm}\includegraphics[scale=0.12]{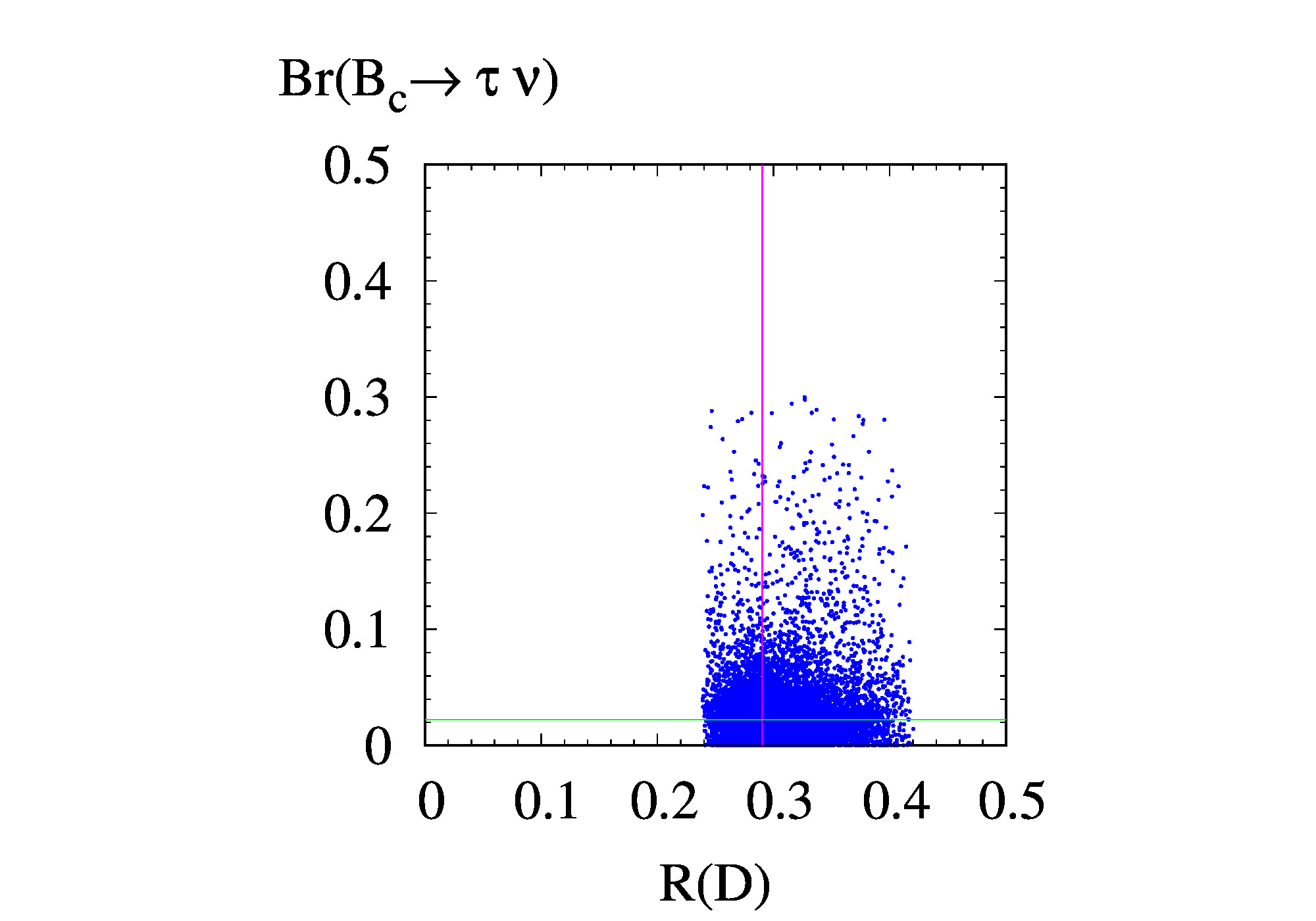} \\
(c) & (d) \\
\hspace{-1cm}\includegraphics[scale=0.12]{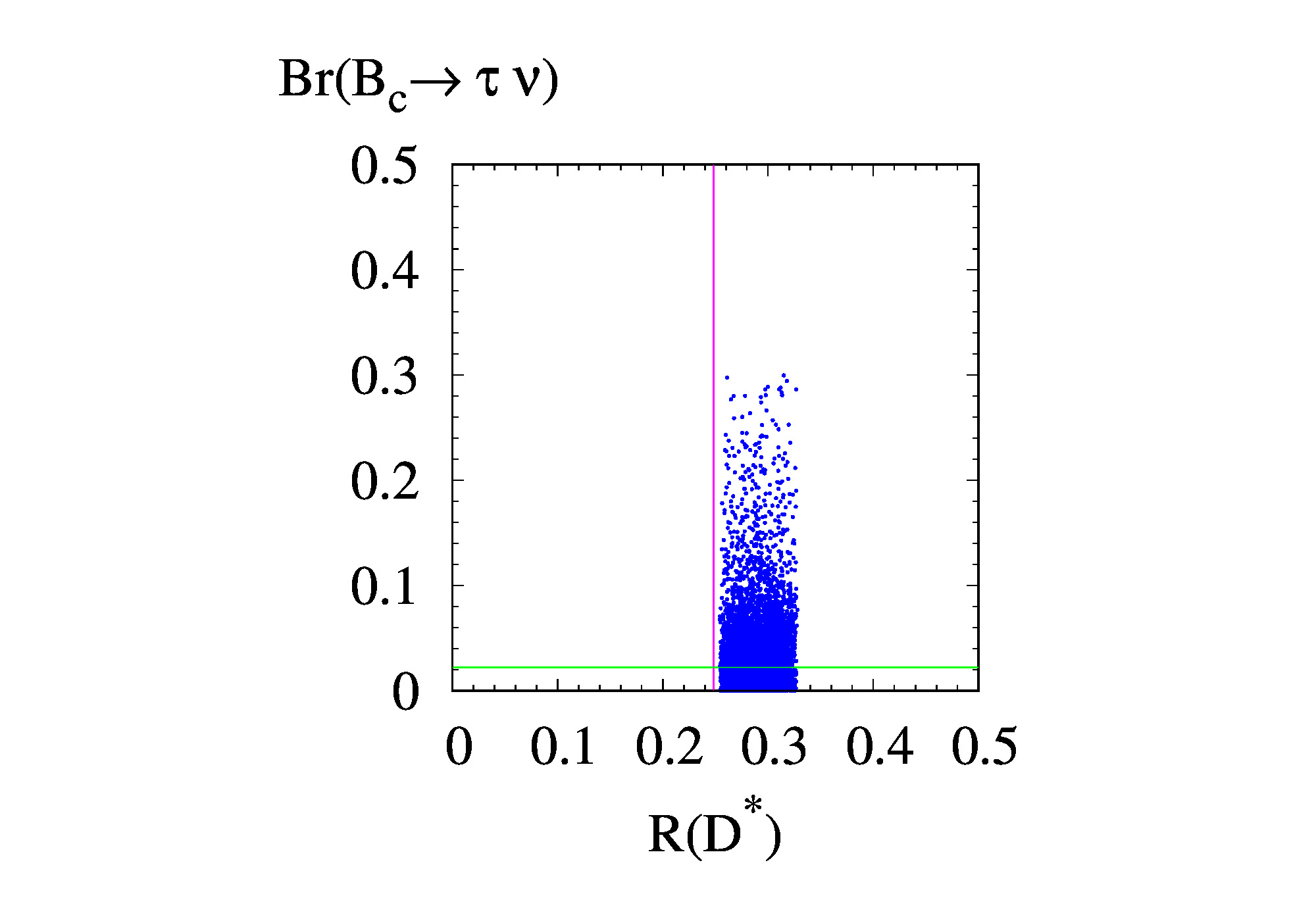} & \\
(e) &
\end{tabular}
\caption{\label{RDRDs} Allowed regions of 
(a) $R(D^*)$ vs. $R(D)$,
(b) $P_\tau(D^*)$ vs. $P_\tau(D)$ with respect to $F_L(D^*)$,
(c) $P_\tau(D^*)$ vs. $P_\tau(D)$ for $C_T\ne 0$ (blue) and $C_T=0$ (red),
(d) $\Br(B_c\to\tau\nu)$ vs.  $R(D)$, and 
(e) $\Br(B_c\to\tau\nu)$ vs. $R(D^*)$ 
at the $2\sigma$ level.
Green and magenta lines are the SM predictions.
}
\end{figure}
%----------------------------------------------------------------------------
%
\par
Figure \ref{RDRDs} shows the allowed regions of some observables.
As shown in Fig. \ref{RDRDs} (a), $R(D^*)$ has no overlaps with the SM. 
Figure \ref{RDRDs} (b) provides allowed regions of polarization parameters
$P_\tau(\Ds)$ and $F_L(D^*)$.
As discussed before $P_\tau(D)$ is mostly positive but negative values are also possible.
Figure \ref{RDRDs} (c) shows the subregion of $C_T=0$ (red points).
The NP scenario of $C_T=0$ with $C_{VL,SR}\ne 0$ is very marginal. 
Figures \ref{RDRDs} (d) and (e) plot the branching ratio $\Br(B_c\to\tau\nu)$ vs. $R(\Ds)$.

%
%
%----------------- Figure 8 : For R(J/Psi) & R(Lambda_c)  ------------------------------------------------
\begin{figure}
\begin{tabular}{cc}
\hspace{-1cm}\includegraphics[scale=0.12]{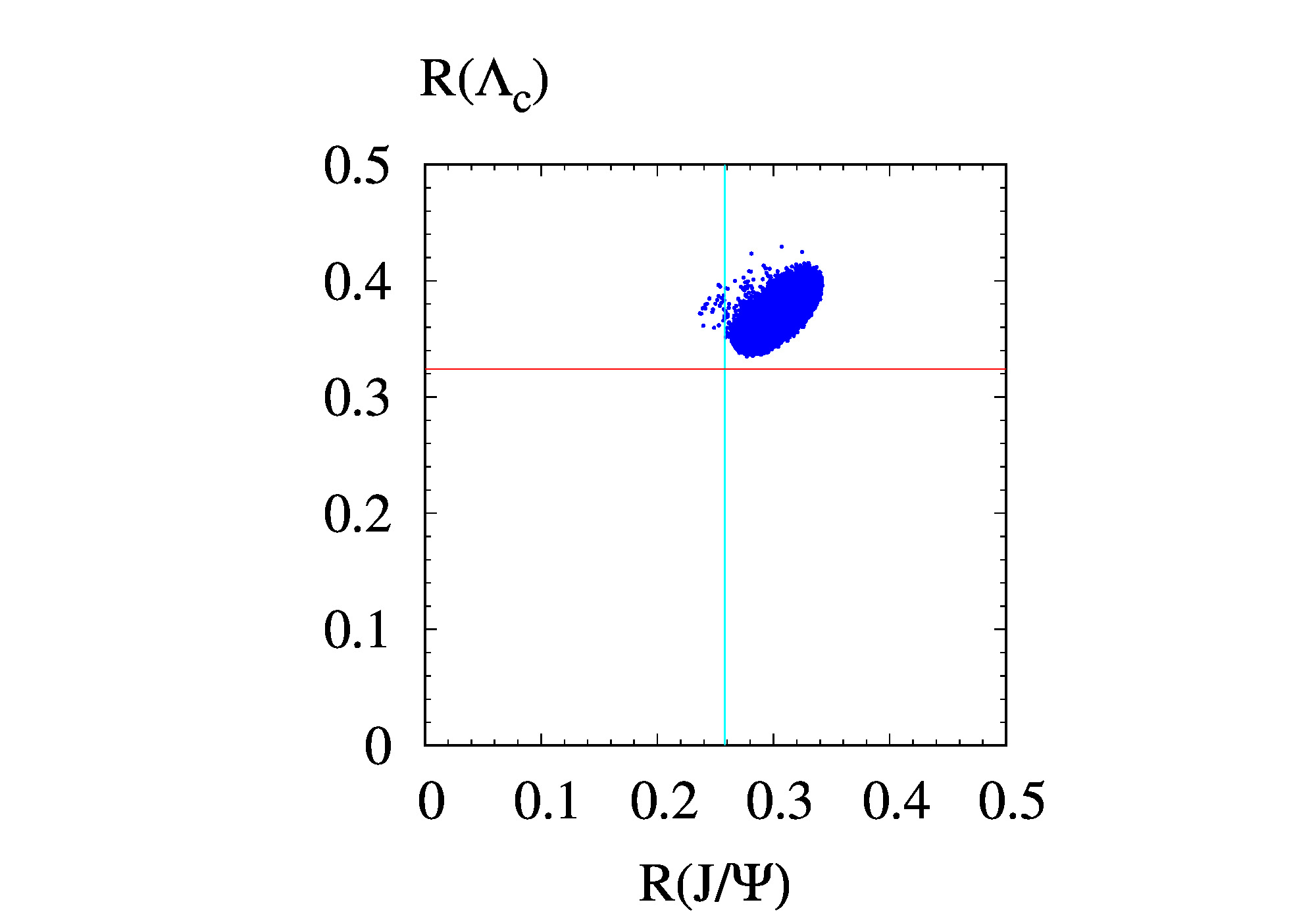} &
\hspace{-1cm}\includegraphics[scale=0.12]{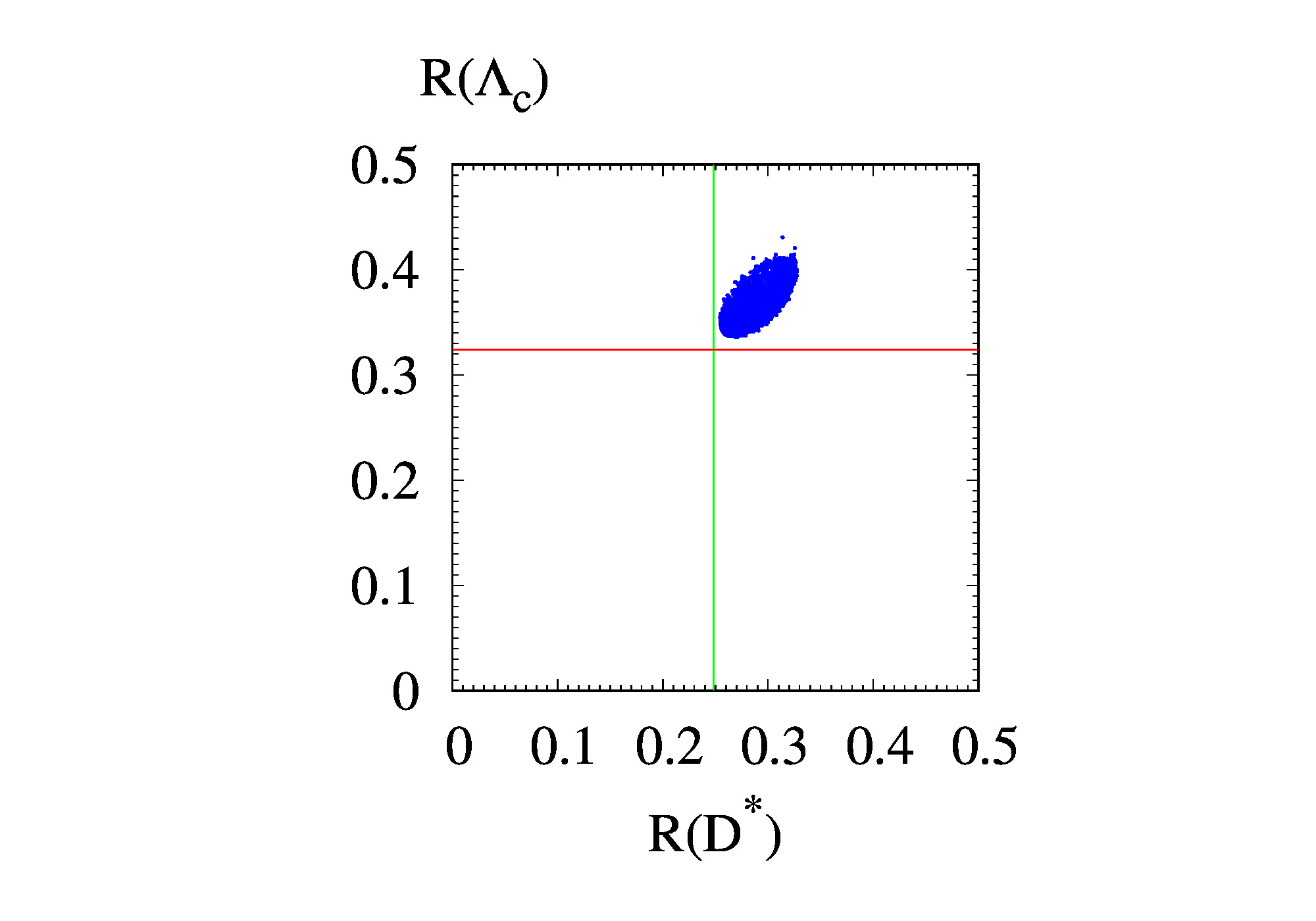} \\
(a) & (b) \\
\hspace{-1cm}\includegraphics[scale=0.12]{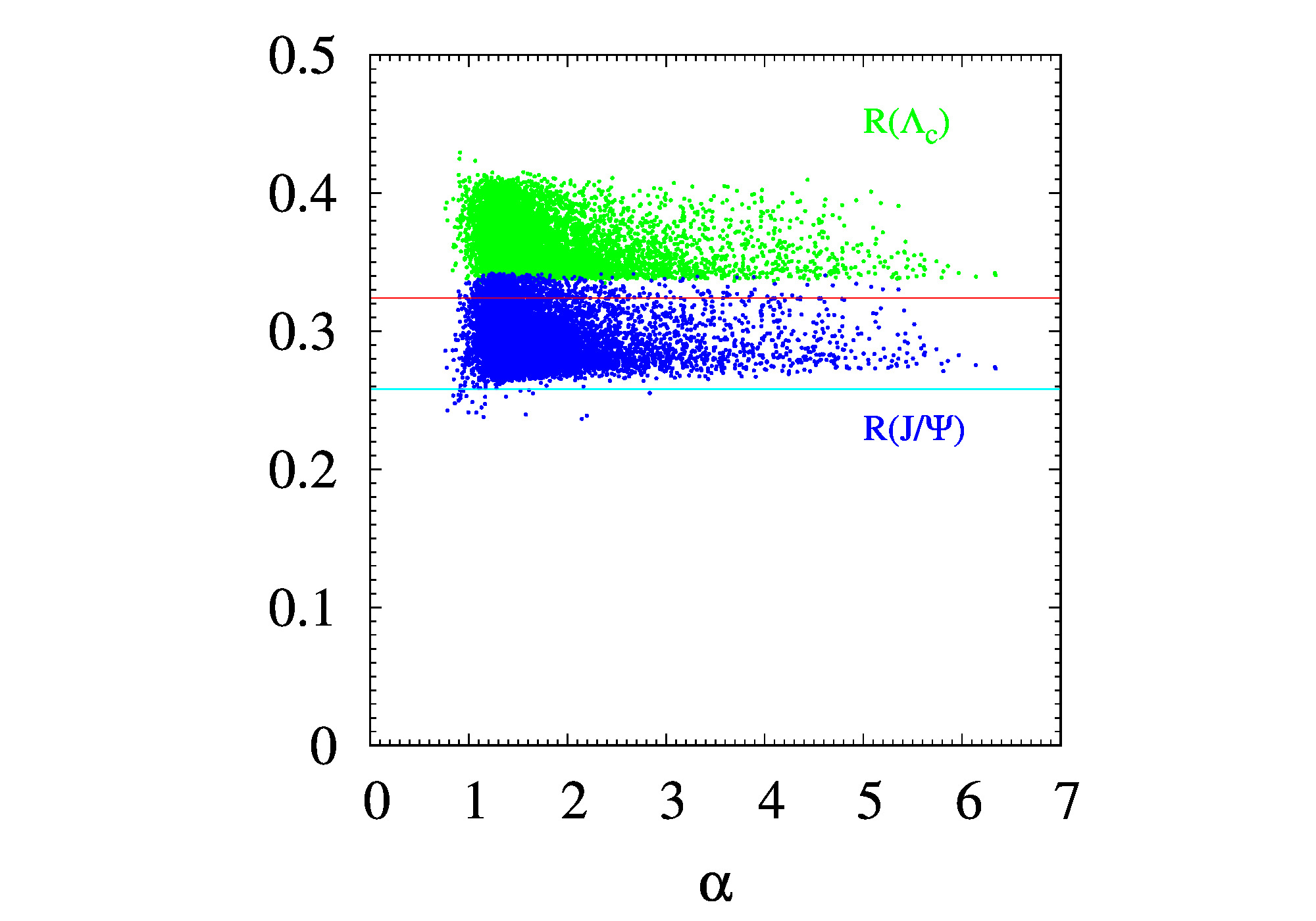} &
\hspace{-1cm}\includegraphics[scale=0.12]{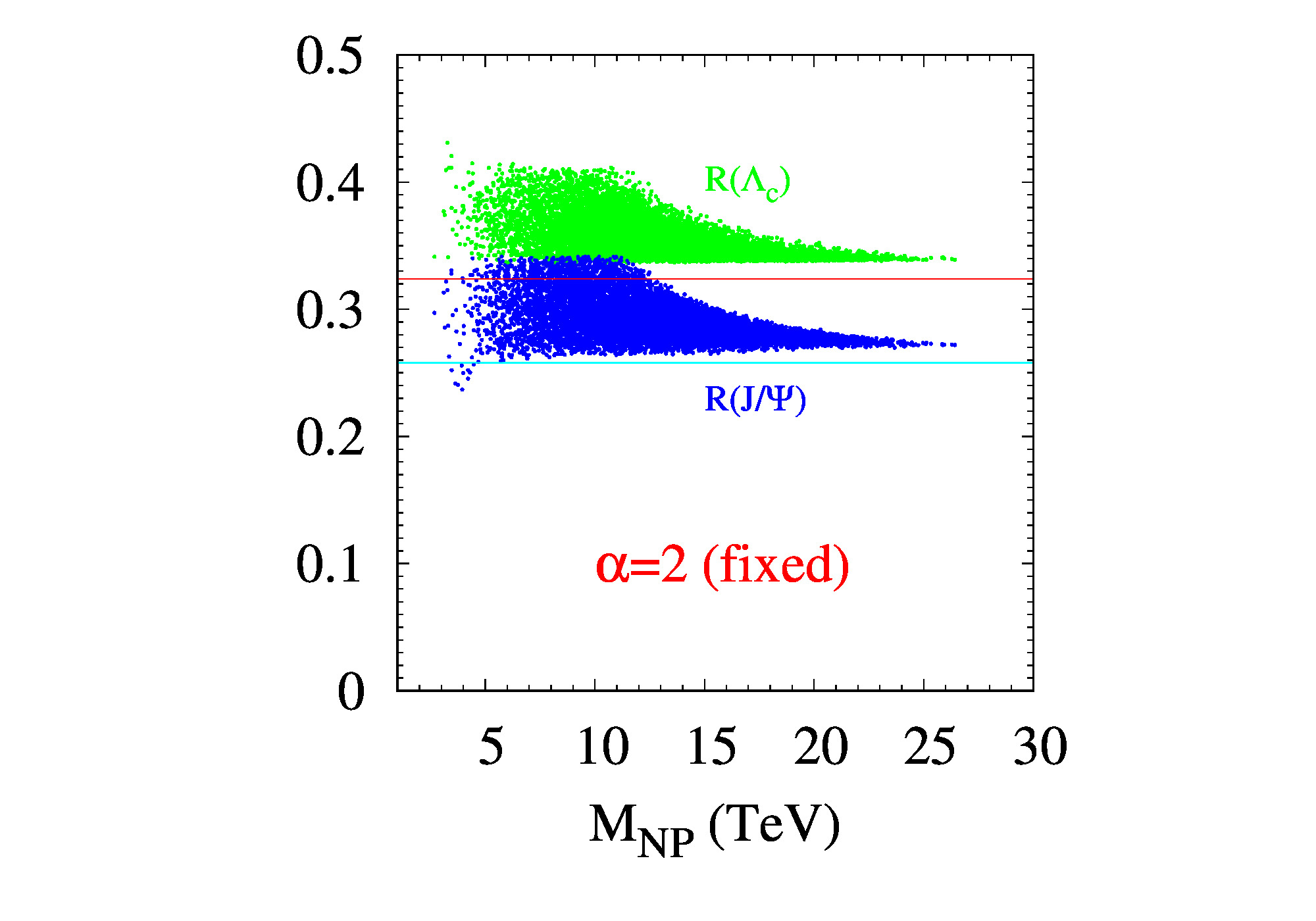} \\ 
(c) & (d) \\ 
\hspace{-1cm}\includegraphics[scale=0.12]{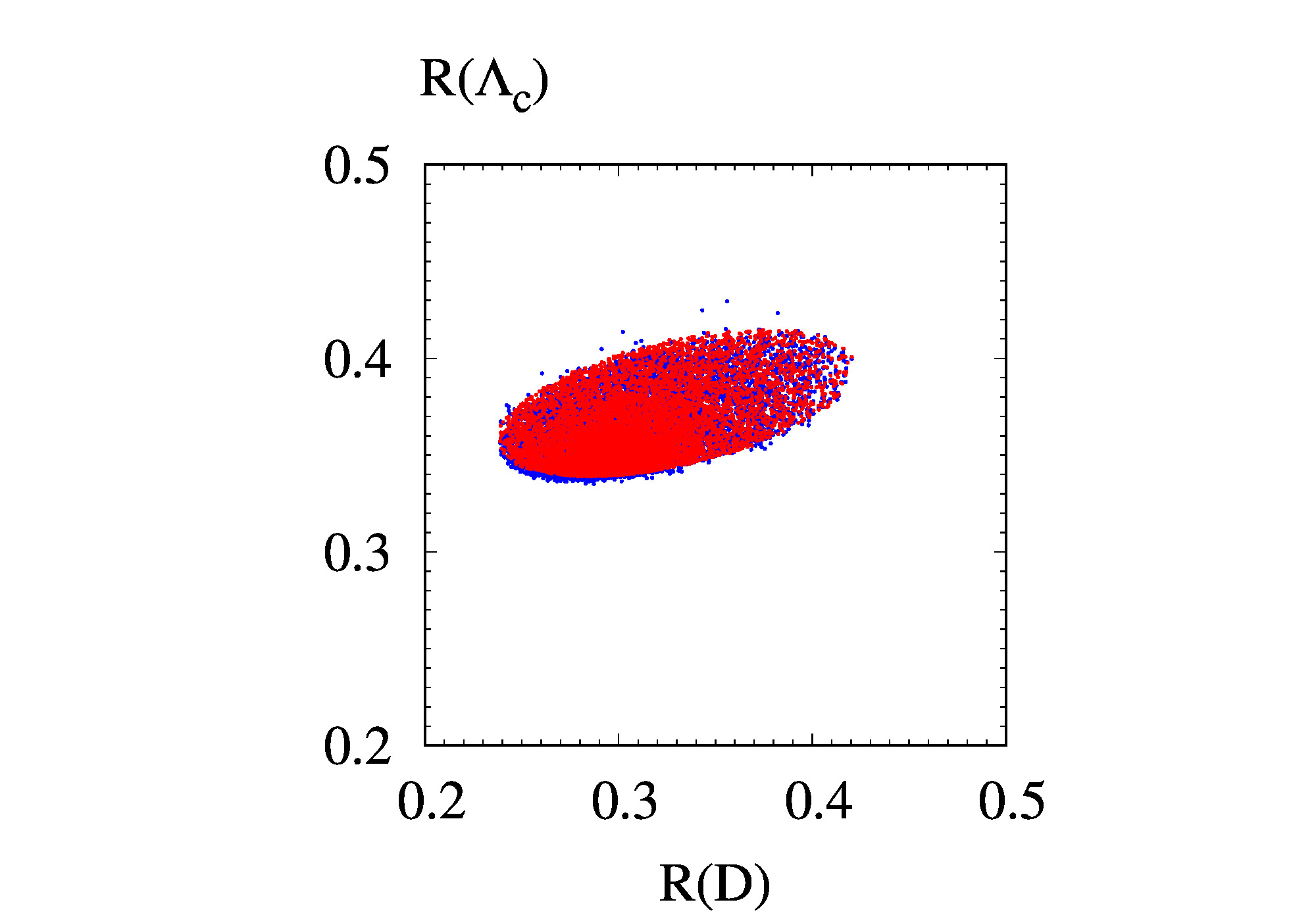} &
\hspace{-1cm}\includegraphics[scale=0.12]{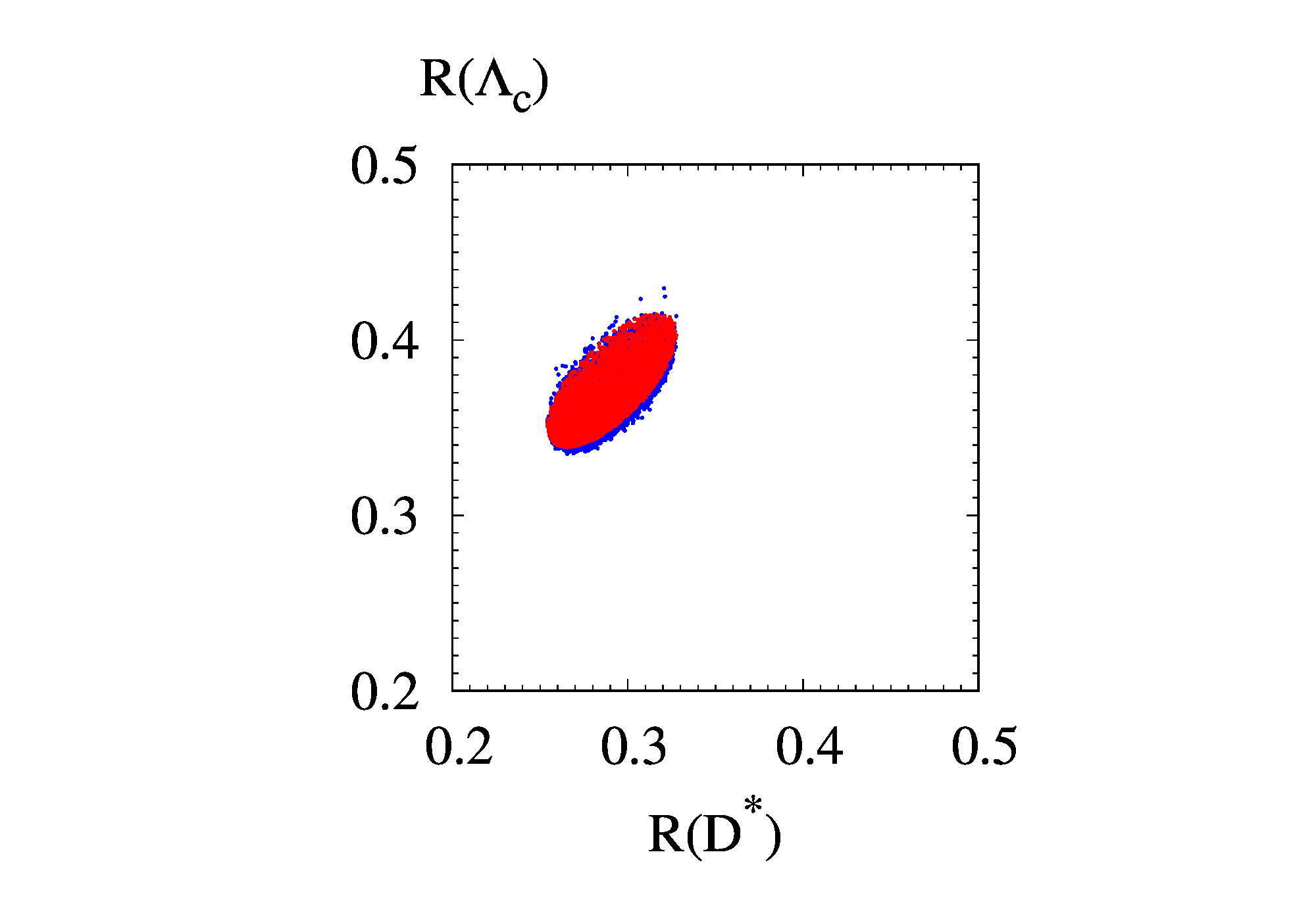} \\ 
(e) & (f) \\ 
\end{tabular}
\caption{\label{RJPRLdc} Allowed regions of 
(a) $R(\Lambda_c)$ vs. $R(J/\Psi)$, 
(b) $R(\Lambda_c)$ vs. $R(D^*)$
(c) $R(J/\Psi)$ and $R(\Lambda_c)$ vs. $\alpha$,
(d) $R(J/\Psi)$ and $R(\Lambda_c)$ vs. $M_\NP$ for fixed $\alpha=2$,
(e) $R(\Lambda_c)$ vs. $R(D)$ from Eq.\ (\ref{RLdcnum}) (blue) and 
from the sum rule of Eq.\ (\ref{sumrule}), and
(f) $R(\Lambda_c)$ vs. $R(D^*)$ from Eq.\ (\ref{RLdcnum}) (blue) and 
from the sum rule of Eq.\ (\ref{sumrule}),
at the $2\sigma$ level.
Straight lines are the SM predictions for 
$R(J/\Psi)$ (cyan), $R(\Lambda_c)$ (red), and $R(D^*)$ (green)
.}
\end{figure}
%------------------------------------------------------------------------------------------------------------------------
%
\par
Figure \ref{RJPRLdc} deals with $R(J/\Psi)$ and $R(\Lambda_c)$.
The best-fit values of our analysis are 
$R(J/\Psi)_{\rm best-fit}=0.303$ and $R(\Lambda_c)_{\rm best-fit}=0.379$.
The former is smaller than the experimental data while the latter larger, 
as the SM predictions.
Our allowed regions are mostly deviated from the SM predictions, and
$R(\Lambda_c)$ has no overlaps with the SM.
Figure \ref{RJPRLdc} (b) plots $R(\Lambda_c)$ vs. $R(D^*)$.
Similar to previous observables, $R(J/\Psi)$ and $R(\Lambda_c)$ tends to get closer to the SM predictions
as $\alpha$ and $M_\NP$ (with fixed $\alpha=2$) gets larger as shown in Figs.\ \ref{RJPRLdc} (c) and (d).
Both of these two ratios have values in the vast majority slightly above the SM predictions.
In Figs.\ \ref{RJPRLdc} (e) and (f) we check the validity of the sum rule 
for $R(\Lambda_c)$ in Eq.\ (\ref{sumrule}).
Blue dots are from Eq.\ (\ref{RLdcnum}) while red ones from the sum rule.
One can find that the sum rule works quite well.
The reason is that the allowed range of $|C_T|$ is much smaller than 1, 
and it ensures the validity of the sum rule as discussed in \cite{Iguro2405}.
More data would check this feature and the consistency of the whole description for $b\to c$ 
semi-leptonic decays.
\par
%%%%%%%%%%%%%%%%%%%%%%%%%%%
%
%\textcolor{red}{$<<$ISSUE2$>>$}
%
%%%%%%%%%%%%%%%%%%%%%%%%%%%
Final remarks are in order to discuss the theoretical implications of the parameter constraints.
Our analysis can be used to check the validity of certain NP scenarios in a generic way 
with respect to $b\to c\tau\nu$ decays.
As an example, for a weak doublet scalar LQ $R_2$, one has 
$A_{SL}^{R_2}=y_{2~c\tau}^{LR}(y_{2~b\tau}^{RL})^*/(2V_{cb})$ \cite{DAlise2403} 
where $y_{2~ij}^{AB}$ are the relevant couplings.
For order one couplings the mass of $R_2$ is required to go over 20 TeV by the $b\to s$ data.
Similar analysis shows that the mass of the vector weak doublet LQ $V_2$ 
should be larger than 30 TeV \cite{DAlise2403}.
This is quite challenging because our analysis requires $M_\NP\lesssim 27~\TeV$ for $|A_j|\le 100$.
On the other hand, for the scalar weak triplet LQ $S_3$ and the vector weak triplet LQ $U_3$ 
with order one couplings, the lower bound of the LQ scale is about 50 TeV by $b\to s$ data \cite{DAlise2403}.
According to our analysis, $S_3$ and $U_3$ are not compatible with $b\to c$ transition.
For other NP models one can do the similar estimations to support or disfavor the model.
\par
In addition, as discussed before, negative $P_\tau(D)$ if observed would strongly constrain 
the Lorentz structure of NP such that $C_{VL, SR}\ne 0$ or $C_{VL,SL,T}\ne 0$.
The last term of $P_\tau(D)$ in Eq.\ (\ref{PtD}), $-1.09(1+C_{VL})C_T^*$ plays an important role,
and negatively large $P_\tau(D)$ requires $C_T\ne 0$.
%
%
%
%
%
%
%
%%%%%%%%%%%%%%%%%%%%%%%%%%%%%%%%%%%%%%%%%%%%%%%%%%%%%%
\section{Conclusions}
%%%%%%%%%%%%%%%%%%%%%%%%%%%%%%%%%%%%%%%%%%%%%%%%%%%%%%
%
In conclusion we investigated possible NP effects at tree level on $B\to\Ds\ell\nu$ decays
with operators $\calO_{VL,SL,SR,T}$.
Corresponding Wilson coefficients $C_j$ are parameterized by a new NP scale $M_\NP$
and its power with fermionic couplings $A_j$.
Allowed regions of $|C_T|$ are narrower than other Wilson coefficients $|C_{VL,SL,SR}|$.
The analysis was done for  $-100\le A_j\le 100$.
Within this range we found that the NP scale is $M_\NP\lesssim 27~\TeV$ for ordinary new particles
where $\alpha=2$.
One should keep in mind that if $|A_j|$ could be larger, then 
bounds on $M_\NP$ discussed above can also be larger accordingly. 
%
%.%%%%%%%%    Newly updated
%
Conversely, current accelerator energy is quite below 27 TeV so that only the region of small $|A_j|$
could be investigated experimentally.
%
%%%%%%%%%%%%%%%%%%%%%%%
\par
The ratio $R(D^*)$ turned out to be still far away from the SM.
In our framework the polarization asymmetry $P_\tau(D)$ can have negative values
while the SM prediction of it is positive.
We found that negative $P_\tau(D)$ is possible for $\calO_{VL,SR}\ne 0$ or 
$\calO_{VL,SL,T}\ne 0$, and
models with $(\calO_{VL,SR}\ne 0, \calO_T=0)$ is marginal. 
If $P_\tau(D)$ turned out to be negatively large, then $C_T$ should be nonzero.
Also, negative $P_\tau(D)$ would impose a strong bound on the NP scale as
$M_\NP \lesssim 6~\TeV$ for $\alpha=2$.
Thus the measurement of $P_\tau(D)$ would be very interesting in probing the NP structure.
\par
In our analysis $R(J/\Psi)$ has slight overlaps with SM while $R(\Lambda_c)$ does not.
We checked that the sum rule for $R(\Lambda_c)$ with respect to $R(\Ds)$ works very well
thanks to small values of $|C_T|$.
\par
In specific models, one should consider many other constraints from processes such as 
$B\to DD$ decay, $B_s\to\mu\mu$ decay, $B_{(s)} \Bbar_{(s)}$ mixing, and so on.
The results would directly affect $A_j$ and possibly $\alpha$.
Our framework could provide more general ways to handle various kinds of NP scenarios,
%.%%%%%%%%    Newly updated
and check whether we would support or exclude certain NP models.
%
%%%%%%%%%%%%%%%%%%%%%%%

%
%
%
%
%
%
%%%%%%%%%%%%%%%%%%%%%%%%%%%%%%%%%%%%%%%%%%%%%%%%%%%%%%
\begin{acknowledgments}
This paper was supported by Konkuk University in 2024.
\end{acknowledgments}
%%%%%%%%%%%%%%%%%%%%%%%%%%%%%%%%%%%%%%%%%%%%%%%%%%%%%%
%
%
%
%
%
%%%%%%%%%%%%%%%%%%%%%%%%%%%%%%%%%%%%%%%%%%%%%%%%%%%%%%

\end{document}